\pdfoutput=1
\documentclass[a4paper,12pt]{article}

\usepackage[pdftex,breaklinks,colorlinks]{hyperref}
\hypersetup{  
  pdftitle =
    {Modified Ehrenfest-TDDFT dynamics},
  pdfauthor =
    {Pablo Echenique},
  pdfsubject=
    ab initio molecular dynamics,
  pdfkeywords =
    Ehrenfest TDDFT ab initio molecular dynamics Born-Oppenheimer non-adiabatic Hellmann-Feynman,
}
\hypersetup{  
    citecolor = blue,
    urlcolor = blue
}

\usepackage{amstext,amsmath,amssymb} 
\usepackage{txfonts}                 
\usepackage{cancel}                  

\usepackage{graphicx}
\usepackage{epsfig}
\usepackage[]{color}

\usepackage[square,comma,numbers,sort&compress]{natbib}
\usepackage{hypernat}  

\usepackage{vmargin} 

\begin{document}

\setcounter{secnumdepth}{2}  
\numberwithin{equation}{section}
\numberwithin{figure}{section}

\title{A modified Ehrenfest formalism for efficient\\
       large-scale ab initio molecular dynamics}

\author{Xavier Andrade$^1$\footnote{Corresponding author. Email: {\tt xavier@tddft.org}}\ , Alberto Castro$^2$, David Zueco$^{3,4}$, J. L. Alonso$^{4,5}$,\\ Pablo Echenique$^{4,5}$, Fernando Falceto$^{4,5}$ and Angel Rubio$^1$\footnote{Corresponding author. Email: {\tt angel.rubio@ehu.es}}
\vspace{0.4cm}\\
{\small $^{1}$ Nano-bio Spectroscopy Group and European Theoretical Spectroscopy Facility,}\\[-4pt]
{\small Departamento de F{\'{\i}}sica de Materiales, Universidad del Pa{\'{i}}s Vasco,}\\[-4pt]
{\small Centro Mixto CSIC-UPV, and DIPC,}\\[-4pt] 
  {\small Edificio Korta, Av. Tolosa 72, E-20018 San Sebasti\'an, Spain.}\\
{\small $^{2}$ Institut f\"ur Theoretisch Physik, Freie Universit\"at Berlin,}\\[-4pt] 
  {\small Arnimallee, 14, Berlin 14195, Deutschland.}\\
{\small $^{3}$ Institut f\"ur Physik, Universit\"at Augsburg,}\\[-4pt] 
  {\small Universit\"atsstra\ss e 1, D-86135 Augsburg, Germany.}\\
{\small $^{4}$ Instituto de Biocomputaci\'on y F\'{\i}sica de los Sistemas Complejos (BIFI),}\\[-4pt] 
  {\small Edificio Cervantes, Corona de Arag\'on 42, E-50009, Zaragoza, Spain.}\\
{\small $^{5}$ Departamento de F\'{\i}sica Te\'orica, Universidad de Zaragoza,}\\[-4pt]
  {\small Pedro Cerbuna 12, E-50009, Zaragoza, Spain.}}

\date{\today}

\maketitle

\begin{abstract}
  We present in detail the recently derived ab-initio molecular
  dynamics (AIMD) formalism [Phys. Rev. Lett. {\bf 101} 096403
  (2008)], which due to its numerical properties, is ideal for
  simulating the dynamics of systems containing thousands of atoms.  A
  major drawback of traditional AIMD methods is the necessity to
  enforce the orthogonalization of the wave-functions, which can
  become the bottleneck for very large systems. Alternatively, one can
  handle the electron-ion dynamics within the Ehrenfest scheme where
  no explicit orthogonalization is necessary, however the time step is
  too small for practical applications. Here we preserve the desirable
  properties of Ehrenfest in a new scheme that allows for a
  considerable increase of the time step while keeping the system
  close to the Born-Oppenheimer surface. We show that the
  automatically enforced orthogonalization is of fundamental
  importance for large systems because not only it improves the
  scaling of the approach with the system size but it also allows for
  an additional very efficient parallelization level. In this work we
  provide the formal details of the new method, describe its
  implementation and present some applications to some test
  systems. Comparisons with the widely used Car-Parrinello molecular
  dynamics method are made, showing that the new approach is
  advantageous above a certain number of atoms in the system. The
  method is not tied to a particular wave-function representation,
  making it suitable for inclusion in any AIMD software package.
\end{abstract}

\section{Introduction}
\label{sec:Introduction}

In the last decades the concept of theoretical atomistic simulations
of complex structures in different fields of research (from material
science in general, to biology) has emerged as a third discipline
between theory and experiment. Computational science is now an
essential adjunct to laboratory experiments, it provides
high-resolution simulations that can guide research and serve as tools
for discovery. Today, computer simulations unifying electronic
structure and ion dynamics have come of age, although important
challenges remain to be solved. This ``virtual lab'' can provide
valuable information about complex materials with refined resolution
in space and time, allowing researchers to gain understanding about
the microscopic and physical origins of materials behavior: from
low-dimensional nano-structures, to geology, atmospheric science,
renewable energy, (nano)electronic devices,
(supra)molecular chemistry, etc. Since the numerical
approaches to handle those problems require ``large-scale
calculations'' the success of this avenue of research was only
possible due to the development of high-performance
computers\footnote{As stated by Dirac in 1929, ``The fundamental laws
  necessary for the mathematical treatment of a large part of physics
  and the whole of chemistry are thus completely known, and the
  difficulties lies only in the fact that application of these laws
  leads to equations that are too complex to be solved''}. The present
work addresses our recent developments in the field of
first-principles molecular dynamics simulations. Before getting into
the details, we would like to frame properly the work from a
historical perspective.

Molecular dynamics (MD)\cite{Ciccotti} consists of following the
dynamics of a system of atoms or molecules governed by some
interaction potential; in order to do so, ``one could at any instant
calculate the force on each particle by considering the influence of
each of its neighbors. The trajectories could then be traced by
allowing the particles to move under a constant force for a short-time
interval and then by recalculating a new force to apply for the next
short-time interval, and so on.'' This description was given in 1959
by Alder and Wainwright\cite{Alder1959JCP} in one the first reports of
such a computer-aided calculation\footnote{ Computer simulations of the
  dynamics of systems of interacting molecules based on the Monte
  Carlo methods were presented some years
  before\cite{Metropolis1953JCP}. Also, before the work of Alder and
  Wainwright, some previous "computations" were reported that did not
  utilize modern computers, but rather real physical models of the
  system, i.e. rubber balls linked by rods\cite{Bernal1959Nature}. The
  rapid improvement of digital computer machines discouraged this
  cumbersome, yet entertaining, methodology.}, though the first MD
simulation was probably done by Fermi et al.\cite{Fermi,Ford1992PR}
for a one-dimensional model solid. We can still use this description
to broadly define the scope of MD, although many variants and
ground-breaking developments have appeared during these fifty years,
addressing mainly two key issues: the limitation in the number of
particles and the time ranges that can be addressed, and the accuracy
of the interaction potential.


The former issue was already properly stated by Alder and Wainwright:
``The essential limitations of the method are due to the relatively
small number of particles that can be handled. The size of the system
of molecules is limited by the memory capacity of the computing
machines.'' This statement is not obsolete, although the expression
``small number of particles'' has today of course a very different
meaning -- linked as it is to the exponentially growing capacities of
computers.

The latter issue -- the manner in which the atomic interaction
potential is described -- has also developed significantly over the
years.  Alder and Wainwright used solid impenetrable spheres in the
place of atoms; nowadays, in the realm of the so-called ``classical''
MD one makes use of \emph{force fields}: simple mathematical formulae
are used to describe atomic interactions; the expressions are
parameterized by fitting either to reference first-principles
calculations or experimental data. These models have become extremely
sophisticated and successful, although they are ultimately bound by a
number of limitations. For example, it is difficult to tackle
electronic polarization effects and one needs to make use of
polarizable models, whose transferability is very questionable but are
widely used with success in many situations.  Likewise, the force
field models are constructed assuming a predetermined bond
arrangement, disabling the option of chemical reactions -- some
techniques exist that attempt to overcome this
restriction\cite{War1980JACSa}, but they are also difficult to
transfer and must be carefully adapted to each particular system.

The road towards precise, non-empirical inter-atomic potentials
reached its destination when the possibility of performing \emph{ab
  initio} MD (AIMD) was realized\cite{Mar2000TR,Tuc2002JPCM}.  In this
approach, the potential is not modelled a priori via some
parameterized expression, but rather generated ``on the fly'' by
performing accurate first-principles electronic structure
calculations. The accuracy of the calculation is therefore limited by
the level of theory used to obtain the electronic structure --
although one must not forget that underlying all MD simulations is the
electronic-nuclear separation \emph{ansatz}, and the classical limit
for the nuclei. The use of very accurate first principles methods for
the electrons implies very large computational times, and therefore it
is not surprising that AIMD was not really born until
density-functional theory (DFT) became mature -- since it provides the
necessary balance between accuracy and computational
feasibility\cite{PrimerDFT,HohenbergKohn,KohnSham,KohnNobel}. Of
fundamental importance was the development of gradient generalized
exchange and correlation functionals, like the ones proposed by
John~Perdew\cite{Per1986PRBa,Per1986PRBb,Per1996PRL}, that can
reproduce experimental results better than the local density
approximation \cite{Car1996Book,Laa1993JCP,Fav1999PRB}. In fact, the
whole field of AIMD was initiated by Car and Parrinello in
1985\cite{Car1985PRL}, in a ground-breaking work that unified DFT and
MD, and introduced a very ingenious acceleration scheme based on a
fake electronic dynamics. As a consequence, the term AIMD in most
occasions refers exclusively to this technique proposed by Car and
Parrinello. However, it can be understood in a more general sense,
including more possibilities that have developed thereafter -- and in
the present work we will in fact discuss one of them. The new scheme
proposed below will benefit from all the algorithm developments and
progress being done in the CP framework.

As a matter of fact, the most obvious way to perform AIMD would be to
compute the forces on the nuclei by performing electronic structure
calculations on the ground-state Born-Oppenheimer potential energy
surface. This we can call ground-state Born Oppenheimer MD
(gsBOMD). It implies a demanding electronic minimization at each step
and schemes using time-reversible integrators have been recently
developed\cite{Nik2008PRL,Nik2006PRL}.  The Car-Parrinello (CP)
technique is a scheme that allows to propagate the Kohn-Sham (KS)
orbitals with a fictitious dynamics that nevertheless mimics gsBOMD --
bypassing the need for the expensive minimization. This idea has
produced an enormous impact, allowing for successful applications in a
surprisingly wide range of areas (see the special number in
ref.~\citealp{And2005CPC} and references therein). Still, it
implies a substantial cost, and many interesting potential
applications have been frustrated due to the impossibility of
attaining the necessary system size or simulation time length.  There
have been several efforts to refine or redefine the CP scheme in order
to enhance its power: linear scaling methods\cite{Goe1999RMPa} attempt
to speed-up in general any electronic structure calculation; the use
of a localized orbital representation (instead of the much more common
plane-waves utilized by CP practitioners) has also been
proposed\cite{Sch2001JCP}; recently, K{\"{u}}hne and
coworkers\cite{Kuh2007PRLa} have proposed an approach which is based
on CP, but which allows for sizable gains in efficiency. In any case,
the cost associated with the orbital orthonormalization that is
required in any CP-like procedure is a potential bottleneck that
hinders its application to very large-scale simulations.

Another possible AIMD strategy is Ehrenfest MD, to be presented in the
following section. In this case, the electron-nuclei separation
\emph{ansatz} and the
Wentzel-Kramers-Brillouin\cite{Wen1926ZPa,Kra1926ZPa,Bri1926CRa}
(WKB) classical limit are also considered; however, the electronic
subsystem is not assumed to evolve on only one of the electronic
adiabatic states -- typically the ground-state one. Instead the
electrons are allowed to evolve on an arbitrary wave-function that
corresponds to a combinations of adiabatic states. As a drawback, the
time-step required for a simulation in this scheme is determined by
the maximum electronic frequencies, which means about three orders of
magnitude less than the time step required to follow the nuclei in a
BOMD.

If one wants to do Ehrenfest-MD, the traditional ``ground-state'' DFT
is not enough, and one must rely on time-dependent density-functional
theory (TDDFT)\cite{Mar2006, Gro1984PRLa}. Coupling TDDFT to
Ehrenfest MD provides with an orthogonalization-free alternative to CP
AIMD -- plus it allows for excited-states AIMD. If the system is such
that the gap between the ground-state and the excited states is large,
Ehrenfest-MD tends to gsBOMD. The advantage provided by the lack of
need of orthogonalization is unfortunately offset by the smallness of
the required time-step\cite{Car1996Book, Mar2000TR}. Recently, some of
the authors of the present article have presented a formalism for
large scale AIMD based on Ehrenfest and TDDFT, that borrows some of
the ideas of CP in order to increase this time step and make
TDDFT-Ehrenfest competitive with CP\cite{Alo2008PRL}.

This article intends to provide a more detailed description of this
proposed methodology: we start, in Section~\ref{sec:Ehrenfest} by
revisiting the mathematical route that leads from the full
many-particle electronic and nuclear Schr{\"{o}}dinger equation to the
Ehrenfest MD model. Next, we clear up some confusions sometimes found
in the literature related to the application of the Hellmann-Feynman
theorem, and we discuss the integration of Ehrenfest dynamics in the
TDDFT framework. Section~\ref{sec:Modified} presents in detail the
aforementioned novel formalism, along with a discussion regarding
symmetries and conservation laws. Sections \ref{sec:Methods} and
\ref{sec:Numerical} are dedicated to the numerical technicalities,
including several application examples.


\section{Ehrenfest dynamics: fundaments and implications for first
  principles simulations}
\label{sec:Ehrenfest}

The starting point is the time-dependent Schr{\"o}dinger equation
(atomic units \cite{Ech2007MP} are used throughout this paper) for a
molecular system described by the wave-function 
$\Phi\big(\{x_j\}_{j=1}^n,\{X_J\}_{J=1}^N,t\big)$:
\begin{equation}
\label{eq:mol_tdSch}
i\dot{\Phi}\big(\{x_j\}_{j=1}^n,\{X_J\}_{J=1}^N,t\big) =
\hat{H} \Phi\big(\{x_j\}_{j=1}^n,\{X_J\}_{J=1}^N,t\big) \ ,
\end{equation}
where the dot indicates the time derivative, and we denote as $\vec{r}_j$,
$\sigma_j$, and $\vec{R}_J$, $\Sigma_J$ the Euclidean coordinates and the spin
of the $j$-th electron and the $J$-th nuclei, respectively, with
$j=1,\ldots,n$, and $J=1,\ldots,N$. We also define $x_j :=
(\vec{r}_j,\sigma_j)$, and $X_J := (\vec{R}_J,\Sigma_J)$, and we shall denote
the whole sets $r:=\{\vec{r}_j\}_{j=1}^n$, $R:=\{\vec{R}_J\}_{J=1}^N$,
$x:=\{x_j\}_{j=1}^n$, and $X:=\{X_J\}_{J=1}^N$, using single letters in order
to simplify the expressions.

The non-relativistic molecular Hamiltonian operator is defined as
\begin{eqnarray}
\label{eq:mol_Ham}
\hat{H} &:=& -\sum_J \frac{1}{2M_J}\nabla_J^2
             -\sum_j \frac{1}{2}\nabla_j^2
             +\sum_{J<K} \frac{Z_JZ_K}{|\vec{R}_J - \vec{R}_K|}
             +\sum_{j<k} \frac{1}{|\vec{r}_j - \vec{r}_k|}
             -\sum_{J,j} \frac{Z_J}{|\vec{R}_J - \vec{r}_j|} \nonumber \\
        &=:& -\sum_J \frac{1}{2M_J}\nabla_J^2
             -\sum_j \frac{1}{2}\nabla_j^2
             +\hat{V}_\mathrm{n-e}(r,R) \nonumber \\
        &=:& -\sum_J \frac{1}{2M_J}\nabla_J^2
             +\hat{H}_\mathrm{e}(r,R) \ ,
\end{eqnarray}
where all sums must be understood as running over the whole natural
set for each index, unless otherwise specified. $M_J$ is the mass of
the $J$-th nucleus in units of the electron mass, and $Z_J$ is the
charge of the $J$-th nucleus in units of (minus) the electron
charge. Also note that we have defined the nuclei-electrons potential 
$\hat{V}_\mathrm{n-e}(r,R)$ and the electronic Hamiltonian
$\hat{H}_\mathrm{e}(r,R)$ operators.

The initial conditions of eq.~(\ref{eq:mol_tdSch}) are given by
\begin{equation}
\label{eq:mol_ic}
\Phi^0 := \Phi(x,X,0) \ ,
\end{equation}
and we assume that $\Phi(x,X,t)$ vanishes at infinity $\forall t$.

Now, in order to derive the quantum-classical molecular dynamics
(QCMD) known as Ehrenfest molecular dynamics from the above setup, one
starts with a separation ansatz for the wave-function $\Phi(x,X,t)$
between the electrons and the nuclei \cite{Ger1982JCP,Ger1988ACP},
leading to the so-called time-dependent self-consistent-field (TDSCF)
equations \cite{Mar2000TR,Bor1996JCP,Bor1995PREP}. The next step is
to approximate the nuclei as classical point particles via short wave
asymptotics, or WKB approximation
\cite{Mar2000TR,Wen1926ZPa,Kra1926ZPa,Bri1926CRa,Bor1996JCP,Bor1995PREP}. The
resultant Ehrenfest MD scheme is contained in the following system of
coupled differential equations \cite{Bor1996JCP,Bor1995PREP}:
\begin{subequations}
\label{eq:Ehrenfest}
\begin{align}
i\dot{\psi}(x,t) & = \hat{H}_\mathrm{e}(r,R(t)) \psi(x,t) \ ,
\label{eq:Ehrenfest_a} \\
M_J \ddot{\vec{R}}_J(t) & =
- \int \mathrm{d}x\, \psi^*(x,t)
  \left[ \vec{\nabla}_J \hat{H}_\mathrm{e}\big(r;R(t)\big) \right]
  \psi(x,t) \ , \quad J=1,\ldots,N
\label{eq:Ehrenfest_b}
\end{align}
\end{subequations}
where $\psi(x,t)$ is the wave-function of the electrons,
$\vec{R}_J(t)$ are the nuclear trajectories, and we have used
$\mathrm{d}x$ to indicate integration over all spatial electronic
coordinates and summation over all electronic spin degrees of
freedom. Also, a semicolon has been used to separate the $r$ from the
$R(t)$ in the electronic Hamiltonian, in order to stress that only the
latter are actual time-dependent degrees of freedom the system.

The initial conditions in Ehrenfest MD are given by
\begin{subequations}
\begin{align}
\psi^0 & := \psi(x,0) \ ,
\label{eq:Ehr_ic_a} \\
\vec{R}_J^0 & := \vec{R}_J(0)
\ , \quad \dot{\vec{R}}_J^0 := \dot{\vec{R}}_J(0) \ , \quad J=1,\ldots,N \ ,
\label{eq:Ehr_ic_b}
\end{align}
\end{subequations}
and we assume that $\psi(x,t)$ vanishes at infinity $\forall t$.

Also note that, since in this scheme $\{\vec{R}_J,\psi\}$ is a set of
independent variables, we can rewrite eqs.~(\ref{eq:Ehrenfest_b}) as
\begin{equation}
\label{eq:Ehrenfest_b2}
M_J \ddot{\vec{R}}_J(t) =
- \vec{\nabla}_J \int \mathrm{d}x \psi^*(x,t)
  \hat{H}_\mathrm{e}\big(r;R(t)\big) \psi(x,t) \ , \quad J=1,\ldots,N
\end{equation}
a fact which is similar in form, but unrelated to the
Hellmann-Feynman theorem\footnote{ Perhaps the first detailed derivation of the
  so-called Hellmann-Feynman theorem was given in G{\"{u}}ttinger,
  P. \emph{Z. Phys.} {\bf 1931}, \emph{73}, 169-184.  The theorem had
  nevertheless been used before that date.\cite{Wal2005} The
  derivations of Hellmann\cite{Hel1937MISC} and
  Feynman\cite{Fey1939PR}, who named the theorem, came a few years
  afterwards.}. As pointed out
by Tully \cite{Tul1998BOOK}, it is likely that the confusion about
whether eqs.~(\ref{eq:Ehrenfest_b2}) should be used to define the
Ehrenfest MD, or the gradient must be applied to the electronic
Hamiltonian inside the integral, as in~(\ref{eq:Ehrenfest_b}), has
arisen from applications in which $\psi(x,t)$ is expressed as a finite
expansion in the set of adiabatic basis functions, $\eta_a(x;R)$,
defined as the eigenfunctions\footnote{In general, $\{\eta_a(x;R)\}$
  may contain a discrete and a continuous part. However, in this
  manuscript, we will forget the continuous part in order to simplify
  the mathematics.} of $\hat{H}_\mathrm{e}(r;R)$:
\begin{equation}
\label{eq:adiabatic_basis}
\hat{H}_\mathrm{e}(r;R(t))\eta_a(x;R(t)) = E_a(R(t)) \eta_a(x;R(t))\ .
\end{equation}

The use of a precise notation, such as the one introduced in this
section, helps to avoid this kind of confusions. An example of a
misleading notation in this context would consist in writing
$\psi(r,R,t)$ for the electronic wave-function
\cite{Tul1998BOOK,Gro2003JCP}, when, as we have emphasized, there
exists no explicit dependence of $\psi$ on the nuclear positions $R$.

In Ehrenfest MD transitions between electronic adiabatic states are
included. This can be made evident by performing the following change
of coordinates from $\{\psi,R\}$ to $\{c,R^\prime\}$ (with
$c:=\{c_a\}_{a=1}^\infty$):
\begin{subequations}
\label{eq:coords_change}
\begin{align}
\psi(x,t) & = \sum_a c_a(t) \eta_a\big(x;R^\prime(t)\big)
 \ , & \label{eq:coords_change_a} \\
\vec{R}_J(t) & = \vec{R}^\prime_J(t) \ , & \quad J=1,\ldots,N \ ,
\label{eq:coords_change_b}
\end{align}
\end{subequations}
where $\eta_a(x;R)$ are known functions given by~(\ref{eq:adiabatic_basis})
and, even if the transformation between the $R$ and the $R^\prime$ is trivial,
we have used the prime to emphasize that there are two distinct sets of
independent variables: $\{\psi,R\}$ and $\{c,R^\prime\}$. This is very
important if one needs to take partial derivatives, since a partial derivative
with respect to a given variable is only well defined when the independent set
to which that variable belongs is specified\footnote{ In the development of the
  classical formalism of Thermodynamics, this point is crucial.}. For example,
a possible mistake is to assume that, since $c_a$ `is independent of'
$\vec{R}^\prime_J$, and $\vec{R}_J = \vec{R}^\prime_J$, then $c_a$ `is also
independent' of $\vec{R}_J$ and, therefore, the unprimed partial derivative
$\vec{\nabla}_Jc_a$ is zero. The flaw in this reasoning is that the unprimed
partial derivative $\vec{\nabla}_J$ is defined to be performed at constant
$\psi$, and not at constant $c$, since the relevant set of independent
variables is $\{\psi,R\}$. In fact, if we write the inverse transformation
\begin{subequations}
\label{eq:coords_change_inv}
\begin{align}
c_a(t) & = \int \mathrm{d}x \psi^*(x,t) \eta_a\big(x;R(t)\big)
\ , & \quad a=1,\ldots,\infty
 \ , \label{eq:coords_change_inv_a} \\
\vec{R}^\prime_J(t) & = \vec{R}_J(t) \ , & \quad J=1,\ldots,N \ ,
\label{eq:coords_change_inv_b}
\end{align}
\end{subequations}
we can clearly appreciate that, even if it is independent from
\(R^\prime\) by construction, \(c_a\) is neither independent from 
\(\psi\), nor from \(R\)\footnote{
  This mistake is made, for example, in ref.~\citealp{Gro2003JCP}
  when going from their expression~(8) to their expression~(9). Having
  not realized the existence of the two distinct sets of independent
  variables: $\{\psi,R\}$ and $\{c,R^\prime\}$, they treat the
  \(|c_a|^2\) as constants when taking the partial derivative with
  respect to \(q\) (our \(R\)) at the right hand side of~(8), thus
  arriving to~(9), where the cross-terms \(\int \mathrm{d}x
  \eta_a^*\big(x;R^\prime(t)\big)
  \vec{\nabla}_J^\prime\hat{H}_\mathrm{e}\big(r;R^\prime(t)\big)
  \eta_b\big(x;R^\prime(t)\big)\), with \(b \ne a\), are incorrectly
  missing (see the correct derivation of EMD in the adiabatic basis
  below).
}
. On the other hand, if we truncated the sum
in~(\ref{eq:coords_change_a}), then there would appear an explicit
dependence of $\psi$ on $R$ and the state of affairs would be
different, since $\{\psi,R\}$ would no longer be a set of independent
variables. However, in the context of an exact (infinite) expansion
in~(\ref{eq:coords_change_a}), the right hand sides of
eqs.~(\ref{eq:Ehrenfest_b}) and~(\ref{eq:Ehrenfest_b2}) are equal, as
we mentioned before, and we do not have to worry about which one is
more appropriate. It is in this infinite-adiabatic basis situation
that we will now use eq.~(\ref{eq:Ehrenfest_b}) and the expansion
in~(\ref{eq:coords_change_a}) to illustrate the non-adiabatic
character of Ehrenfest MD.

If we perform the change of variables described in
eqs.~(\ref{eq:coords_change}) to the Ehrenfest MD eqs.~(\ref{eq:Ehrenfest}), and we use
that
\begin{equation}
\label{eq:calcs1}
\vec{\nabla}_J\hat{H}_\mathrm{e}(r;R) =
\vec{\nabla}_J^\prime\hat{H}_\mathrm{e}(r;R^\prime) \ ,
\end{equation}
we see that we will have to calculate terms of the form
\begin{equation}
\label{eq:calcs2}
\int \mathrm{d}x\, \eta_a^*\big(x;R^\prime(t)\big)
\vec{\nabla}_J^\prime\hat{H}_\mathrm{e}\big(r;R^\prime(t)\big)
\eta_b\big(x;R^\prime(t)\big) \ ,
\end{equation}
which can be easily extracted from the relation
\begin{equation}
\label{eq:calcs3}
\vec{\nabla}_J^\prime\int \mathrm{d}x\, \eta_a^*\big(x;R^\prime(t)\big)
\hat{H}_\mathrm{e}\big(r;R^\prime(t)\big)
\eta_b\big(x;R^\prime(t)\big) =
\vec{\nabla}_J^\prime E_a\big(R^\prime(t)\big)\delta_{ab} \ .
\end{equation}

In this way, we obtain for the nuclear Ehrenfest MD equation
\begin{eqnarray}
\label{eq:Ehrenfest_basis_b}
M_J \ddot{\vec{R}}^\prime_J(t) & = &
- \sum_a |c_a(t)|^2\vec{\nabla}_J^\prime E_a\big(R^\prime(t)\big)
\nonumber \\
&& \mbox{} - \sum_{a,b} c_a^*(t)c_b(t)\left[ E_a\big(R^\prime(t)\big)
  - E_b\big(R^\prime(t)\big)\right]\vec{d}^{ab}_J\big(R^\prime(t)\big) \ ,
\quad J=1,\ldots,N \ ,
\end{eqnarray}
where the non-adiabatic couplings (NACs) are defined as
\begin{equation}
\label{eq:NACs}
\vec{d}^{ab}_J\big(R^\prime(t)\big) :=
\int \mathrm{d}x\,\eta_a^*\big(x;R^\prime(t)\big)
\vec{\nabla}_J^\prime
\eta_b\big(x;R^\prime(t)\big) \ .
\end{equation}

To obtain the new electronic Ehrenfest MD equation, we perform the change of
variables to~(\ref{eq:Ehrenfest_a}) and we then multiply by
$\eta_b^*\big(x;R^\prime(t)\big)$ the resulting expression and
integrate over the electronic coordinates $x$. Proceeding in this
way, we arrive to
\begin{equation}
\label{eq:Ehrenfest_basis_a}
i\hbar\,\dot{c}_a(t) = E_a\big(R^\prime(t)\big) c_a(t)
- i\hbar \sum_b c_b(t) \left[\sum_J \dot{\vec{R}}^\prime_J(t)
  \cdot \vec{d}^{ab}_J\big(R^\prime(t)\big) \right]  \ .
\end{equation}

In the nuclear eqs.~(\ref{eq:Ehrenfest_basis_b}), we can see that the
term depending on the moduli $|c_a(t)|^2$ directly couples the
population of the adiabatic states to the nuclei trajectories, whereas
interferences between these states are included via the $c_a^*(t)c_{b
  \ne a}(t)$ contributions. Analogously, in the electronic equations
above, the first term represents the typical evolution of the
coefficient of an eigenstate of a Hamiltonian, but, differently from
the full quantum case, in Ehrenfest MD, the second term couples the evolution
of all states with each other's through the velocity of the classical
nuclei and the NACs.


Moreover, Ehrenfest MD is fully (quantum) coherent, since the complex
coefficients $c_a (t)$, are the ones corresponding to the quantum
superposition in the electronic wave function.
A proper theory that treats realistically the electronic process of coherence
and decoherence is of fundamental importance to properly interpret transition
rates and to have control over processes happening at the
attosecond/femtosecond time scales, such as the description of the
optimal-pulse laser (in optimal control theory) that enhances a given channel
in a chemical reaction, the manipulation of qbits in quantum computing
devices, the generation of soft X-rays by high-harmonic generation, or the
energy transfer processes in photosynthetic units
\cite{Lee2008Sci}.

At finite temperature, it is known that Ehrenfest MD
cannot account for the Boltzmann equilibrium population of
the quantum subsystem \cite{Mauri,Kab2004,Parandekar2005}.
The underlying reason of this failure is the mean field approximation
in eq.~(\ref{eq:Ehrenfest_b}) which neglects the nuclei response to
the microscopic fluctuations in the electronic charge density.

In order to address this point, it is important to distinguish between
two different physical situations considered in the literature for
studying equilibrium within Ehrenfest: In ref.~\citealp{Mauri}, a
mixed quantum-classical system is coupled, only via the classical
degrees of freedom, to a classical bath. The dissipative dynamics is
integrated using a Nos\'e thermostat, while the mixed
quantum-classical one is integrated using Ehrenfest.  In
ref.~\citealp{Kab2004,Parandekar2005}, on the other hand, the
classical degrees of freedom are the bath to which the quantum system
is coupled, i.e., the thermalization of a quantum system due to its
``Ehrenfest-like'' coupling to a bath or solvent is discussed.  Only
the first of these two approaches corresponds to the physical problem
we want to deal with, namely, the thermalization of a mixed
quantum-classical molecule in a bath.

Once the aforementioned drawback has been recognized, some authors have
proposed several {\it patches} to ensure the Boltzmann population equilibrium
for the quantum subsystem: In Tully's surface hopping (SH) method
\cite{Tully1990}, the quantum degrees of freedom also follow
eq.~(\ref{eq:Ehrenfest_a}), however, instead of the mean-field dynamics
(\ref{eq:Ehrenfest_b}), the classical degrees of freedom follow a
stochastic-like equation describing jumps between adiabatic states.
Unfortunately, this method does not give in general the desired equilibrium
averages either \cite{Schmidt2008}, and it looses the physical meaning of time
during propagation. 

Another new method by Bastida and collaborators
\cite{Bastida2006,Bas2007JCP}, proposes an ad-hoc modification of the
Ehrenfest equations in order to obtain the correct equilibrium
distribution of a quantum system coupled to a classical bath, i.e.,
the classical degrees of freedom are the solvent for the quantum
system. Their idea can be summarized as follows: Expressing in
(\ref{eq:Ehrenfest_basis_a}) the complex coefficients in polar form
($c_a =\rho_a \exp\{{\rm i} \theta_a \}$), and writing the equations
for the moduli, one obtains $\dot \rho_a = - \sum_b \rho_b
cos(\theta_a - \theta_b ) D_{ab}$, where we have used the compact
notation: $ D_{ab}=\sum_J \dot{\vec{R}}^\prime_J(t) \cdot
\vec{d}^{ab}_J\big(R^\prime(t)\big)$,
cf. eq.~(\ref{eq:Ehrenfest_basis_a}).  Analogous equations are derived
for the phases $\theta_a$.  Written like this, the equations are
formally similar to balance-like equations for the diagonal elements
of the density matrix of the quantum system in the adiabatic basis.
These kinetic equations have been extensively studied in relaxation
processes, and it is known that, to ensure equilibrium, the
coefficients, $D_{ab}$, must fulfill the detailed-balance condition,
i.e., $D_{ab} = {\rm exp}\{-\beta \Delta_{ba}\} D_{ba}$, with
$\Delta_{ba}$ being the energy difference between the $b$ and $a$
states\cite{blum}. The proposal by Bastida et
al.\cite{Bastida2006,Bas2007JCP} proceeds by defining some modified
transition coefficients $\widetilde D_{ab}$, such that detailed
balance is enforced, an thus approaching the Boltzmann equilibrium
population for the quantum system.

Coming back to the situation discussed in ref.~\citealp{Mauri}, i.e.,
where the classical subsystem is not a bath, but a part of the mixed
quantum-classical system coupled to a reservoir, one can go beyond
Ehrenfest and make use of the formalism developed in
refs.~\citealp{Kapral1999, Nielsen2001, Kapral2006}. The
description in these works is not mean-field, and the
quantum-classical dynamics is treated {\it exactly}.  Although its
practical implementation seems cumbersome, it is a path to explore,
with possible modifications, in the near future.

Finally we should mention ref.~\citealp{Prezhdo}, where the complementary
situation to ref.~\citealp{Mauri} is studied, coupling the quantum system
directly to the bath, while the classical degrees of freedom are not
coupled directly to any reservoir. A study of the deviations from
equilibrium in this case is still missing.

\section{Ehrenfest-TDDFT}

TDDFT offers a natural framework to implement Ehrenfest MD. In fact,
starting by an extension \cite{Li1986PRA} of the Runge-Gross theorem
\cite{Run1984PRL} to arbitrary multicomponent systems, one can develop
a TDDFT \cite{Gro1996BOOK} for the combined system of electrons and
nuclei described by~(\ref{eq:mol_tdSch}). Then, after imposing a
classical treatment of nuclear motion, one arrives to an
Ehrenfest-TDDFT dynamics. This scheme can also be generated from the
following Lagrangian \cite{Gro1996BOOK,The1992PRB,Alo2008PRL}:
\begin{eqnarray}
\label{eq:L_E-TDDFT}
L\big[\varphi(t),\dot{\varphi}(t),R(t),\dot{R}(t)\big] & := & \frac{i}{2}
\sum_A\int\mathrm{d}\vec{r}
\big( \varphi_A^*(\vec{r},t)\dot{\varphi}_A(\vec{r},t)
-\dot{\varphi}_A^*(\vec{r},t)\varphi_A(\vec{r},t) \big) \nonumber \\
&& \mbox{} + \sum_J \frac{M_J}{2} \dot{\vec{R}}_J(t) \cdot \dot{\vec{R}}_J(t)
-E_\mathrm{KS}\big[\varphi(t),R(t)\big]  \ ,
\end{eqnarray}
where we have denoted by $\varphi := \{\varphi_A\}_{A=1}^{n/2}$, the whole set of
Kohn-Sham (KS) orbitals of a closed-shell molecule, and
$E_\mathrm{KS}[\varphi,R]$ is the KS energy:
\begin{eqnarray}
\label{eq:E_KS}
E_\mathrm{KS}[\varphi,R] & := & 2 \sum_A\int\mathrm{d}\vec{r}
\varphi_A^*(\vec{r},t)\left(-\frac{\nabla^2}{2}\right)\varphi_A(\vec{r},t)
- \int\mathrm{d}\vec{r} \sum_J
  \left( \frac{Z_J}{|\vec{R}_J(t) - \vec{r}|} \right) \rho(\vec{r},t)
 \nonumber \\
&& \mbox{} + \frac{1}{2}\int\mathrm{d}\vec{r}\mathrm{d}\vec{r}^\prime
\frac{\rho(\vec{r},t)\rho(\vec{r}^\prime,t)}{|\vec{r} - \vec{r}^\prime|}
+ E_\mathrm{XC}\big[\rho(\vec{r},t)\big]
+ \sum_{J<K} \frac{Z_JZ_K}{|\vec{R}_J(t) - \vec{R}_K(t)|} \ ,
\end{eqnarray}
where $E_\mathrm{XC}\big[\rho(\vec{r})\big]$ is the exchange-correlation
energy, and the time-dependent electronic density is defined as
\begin{equation}
\label{eq:density}
\rho(\vec{r},t) := 2 \sum_A \big| \varphi_A(\vec{r},t) \big|^2 \ .
\end{equation}

In the following section, we introduce a modification of the
Ehrenfest-TDDFT dynamics obtained from~(\ref{eq:L_E-TDDFT}) aimed to
the study of situations in which the contribution of the
electronic excited states to the nuclei dynamics is negligible,
i.e., situations in which one is interested in performing
ground-state Born-Oppenheimer molecular dynamics (gsBOMD)
\cite{Mar2000TR}.

\section{Modified Ehrenfest-TDDFT formalism}
\label{sec:Modified}

\subsection{Lagrangian and equations of motion}
\label{sec:Lagrangian}

We now introduce the basic concepts and approximations that define the
new fast Ehrenfest-TDDFT dynamics framework that some of the authors
introduced in ref.~\citealp{Alo2008PRL}. The new scheme can
be obtained from the following Lagrangian
\begin{equation}
\label{eq:new_L}
L[\varphi,\dot{\varphi},R,\dot{R}] := \mu \frac{i}{2}
   \sum_A\int\mathrm{d}\vec{r}\left[\varphi_A^*(\vec{r},t)\dot{\varphi}_A(\vec{r},t) - \dot{\varphi}_A^*(\vec{r},t)\varphi_A(\vec{r},t)\right]
 + \sum_J \frac{M_J}{2} \dot{\vec{R}}_J \cdot \dot{\vec{R}}_J
 - E_\mathrm{KS}[\varphi,R]  \ .
\end{equation}

Note that the major modification with respect to the Ehrenfest-TDDFT
Lagrangian in eq.~(\ref{eq:L_E-TDDFT}) is the presence of a parameter
$\mu$ that introduces a re-scaling of the electronic velocities
(Ehrenfest-TDDFT is recovered when $\mu=1$). The equations of motion
of the new Lagrangian, eq.~(\ref{eq:new_L}), are:
\begin{subequations}
\label{eq:new_EoM}
\begin{align}
  i \,\mu \dot{\varphi}_A(\vec{r},t) &=\frac{\delta E_\mathrm{KS}[\varphi,R]}{\delta \varphi_A^*} =
  -\frac{1}{2}\nabla^2 \varphi_A(\vec{r},t) + v_{\mathrm{eff}}[\varphi,R]
  \varphi_A(\vec{r},t) \ , & \quad A=1,\ldots,\frac{n}{2}\ , \label{eq:new_EoM_a}\\
  M_J \ddot{\vec{R}}_J &= - \vec{\nabla}_J E_\mathrm{KS}[\varphi,R] \ , 
  & \quad J=1,\ldots,N \ , \label{eq:new_EoM_b}
\end{align}
\end{subequations}
where $v_{\mathrm{eff}}$ is the time-dependent KS effective
potential. As we are interested in the adiabatic regime, we will
restrict the exchange and correlation potential to depend only on the
instantaneous density (in general the exchange correlation potential
in TDDFT depends on the density of all previous times, although for
practical calculation this same adiabatic approximation is done).

Compare with the gsBOMD Lagrangian:
\begin{equation}
\label{eq:L_BO}
L_\mathrm{BO}[\varphi,R,\dot{R}] := 
 \sum_J \frac{M_J}{2} \dot{\vec{R}}_J \cdot \dot{\vec{R}}_J
 - E_\mathrm{KS}[\varphi,R]  +
 \sum_{AB}\Lambda^\mathrm{BO}_{AB}\left(\int\mathrm{d}\vec{r}
\varphi_A^*(\vec{r},t)\varphi_B(\vec{r},t) - \delta_{AB}\right) \ ,
\end{equation}
and the corresponding equations of motion:
\begin{subequations}
\label{eq:EoM_BO}
\begin{align}
- \frac{1}{2}\nabla^2 \varphi_A(\vec{r},t) + v_{\mathrm{eff}}[\varphi,R] \varphi_A(\vec{r},t)
  & = \sum_B \Lambda^\mathrm{BO}_{AB} \varphi_B(\vec{r},t) 
  \ , & \quad A=1,\ldots,\frac{n}{2}\ , \label{eq:EoM_BO_a}\\
\int\mathrm{d}\vec{r} \varphi_A^*(\vec{r},t)\varphi_B(\vec{r},t)
  & = \delta_{AB} \ , & \quad A,B=1,\ldots,\frac{n}{2}\ , \label{eq:EoM_BO_b}\\
M_J \ddot{\vec{R}}_J &= - \vec{\nabla}_J E_\mathrm{KS}[\varphi,R] \ , 
  & \quad J=1,\ldots,N \ , \label{eq:EoM_BO_c}
\end{align}
\end{subequations}
where $\Lambda^\mathrm{BO}:=(\Lambda^\mathrm{BO}_{AB})$ is a
Hermitian matrix of time-dependent Lagrange multipliers that ensure
that the orbitals $\varphi$ form an orthonormal set at each instant of
time. The Euler-Lagrange equations corresponding to
$\Lambda^\mathrm{BO}_{AB}$ in~(\ref{eq:EoM_BO_b}) are exactly these
orhonormality constraints, and, together with eq.~(\ref{eq:EoM_BO_a}),
constitute the time-independent KS equations. Therefore, assuming no
metastability issues in the optimization problem, the
orbitals $\varphi$ are completely determined\footnote{ This fact is not
  in contradiction with the above discussion about coordinates
  independence. The difference between Ehrenfest MD and gsBOMD is that in the
  Lagrangian for the latter, the orbitals `velocities' $\dot{\varphi}$
  do not appear, thus generating Eqs.~(\ref{eq:EoM_BO_a})
  and~(\ref{eq:EoM_BO_b}), which can be regarded as \emph{constraints}
  between the $\psi$ and the $R$.} by the nuclear coordinates $R$,
being in fact the BO ground state (gs), $\varphi =
\varphi^\mathrm{gs}(R)$, which allows us to write the equations of motion for gsBOMD
in a much more compact and familiar form:
\begin{equation}
\label{eq:EoM_BO_2}
M_J \ddot{\vec{R}}_J = - \vec{\nabla}_J 
  E_\mathrm{KS}\big[\varphi^\mathrm{gs}(R),R\big] \ , 
 \quad J=1,\ldots,N \ .
\end{equation}

We can also compare the dynamics introduced in eqns.~(\ref{eq:new_EoM})
with CPMD, whose Lagrangian reads
\begin{eqnarray}
\label{eq:L_CP}
L_\mathrm{CP}[\varphi,\dot{\varphi},R,\dot{R}] &:=& 
 \frac{1}{2}\mu_\mathrm{CP}\sum_A\int\mathrm{d}\vec{r}|\dot{\varphi}_A(\vec{r},t)|^2 +
 \sum_J \frac{M_J}{2} \dot{\vec{R}}_J \cdot \dot{\vec{R}}_J \nonumber \\
&& \mbox{} - E_\mathrm{KS}[\varphi,R]  +
 \sum_{AB}\Lambda^\mathrm{CP}_{AB}\left(\int\mathrm{d}\vec{r}
\varphi_A^*(\vec{r},t)\varphi_B(\vec{r},t) - \delta_{AB}\right) \ ,
\end{eqnarray}
and the corresponding equations of motion are
\begin{subequations}
\label{eq:EoM_CP}
\begin{align}
  \mu_\mathrm{CP}\ddot{\varphi}_A(\vec{r},t) & =
  -\frac{1}{2}\nabla^2 \varphi_A(\vec{r},t) + v_{\mathrm{eff}}[\varphi,R]
  + \sum_B \Lambda^\mathrm{CP}_{AB} \varphi_B(\vec{r},t)
  \ , & \quad A=1,\ldots,\frac{n}{2}\ , \label{eq:EoM_CP_a}\\
\int\mathrm{d}\vec{r} \varphi_A^*(\vec{r},t)\varphi_B(\vec{r},t)
  & = \delta_{AB} \ , & \quad A,B=1,\ldots,\frac{n}{2}\ , \label{eq:EoM_CP_b}\\
  M_J \ddot{\vec{R}}_J &= - \vec{\nabla}_J E_\mathrm{KS}[\varphi,R] \ , 
  \quad & J=1,\ldots,N \ , \label{eq:EoM_CP_c}
\end{align}
\end{subequations}
where $\Lambda^\mathrm{CP}:=(\Lambda^\mathrm{CP}_{AB})$ is again a
Hermitian matrix of time-dependent Lagrange multipliers that ensure
the orthonormality of the orbitals $\varphi$, and $\mu_\mathrm{CP}$
is a fictitious electrons `mass' which plays a similar role to the
parameter $\mu$ in our new dynamics.

Before discussing in details the main concepts of the present dynamics
it is worth to state its main advantages and deficiencies (that will
be the topic of discussion in the next sections). When applied to
perform gsBOMD the method can gain a large speed-up over Ehrenfest MD,
it preserves exactly the total energy and the wave-function
orthogonality, and allows for a very efficient parallelization scheme
that requires low communication. However, the speed-up comes at a cost
as it increases the non-adiabatic effects. Also, the method as
discussed above will not work properly for metals and small-gap
systems\footnote{We are exploring how to overcome this limitation by
  mapping the real Hamiltonian into another one that produces the same
  dynamics but not having contributions from the empty states.}.

\subsection{Symmetries and conserved quantities}
\label{sec:Symmetries}

In the following we will study the conserved quantities associated to
the global symmetries of the Lagrangian in eq.~(\ref{eq:new_L}) and we
shall compare them with those of gsBOMD and CPMD. We will also be
interested in a gauge symmetry that is the key to understand the
behaviour of eq.~(\ref{eq:new_L}) in the limit $\mu\to 0$ and its relation
with gsBOMD.

The first symmetry we want to discuss is the time translation
invariance of~(\ref{eq:new_L}). This is easily recognized as $L$ does
not depend explicitly on time. Associated to this invariance there is
a conserved `energy'. Namely, using Noether theorem we have that
\begin{equation}
\label{eq:E_new}
E  = \sum_A\int \mathrm{d}\vec{r} 
\left[\frac{\delta L}{\delta\dot{\varphi}_A(\vec{r},t)} \dot{\varphi}_A(\vec{r},t)
+ \frac{\delta L}{\delta\dot{\varphi}_A^*(\vec{r},t)}\dot{\varphi}_A^*(\vec{r},t)\right]
+ \sum_{J,p} \frac{\partial L}{\partial \dot{R}^p_J}\dot{R}^p_J - L 
 =  \sum_J\frac{1}{2}M_J\dot{R}_J^2+E_{KS}[\varphi,R]
\end{equation}
is constant under the dynamics given by eq.~(\ref{eq:new_EoM}), where
$p=1,2,3$ indexes the Euclidean coordinates of vectors
$\dot{\vec{R}}_J$ (and $\vec{R}_J$ if needed).

Notice that $E$ does not depend on the unphysical parameter $\mu$ and
actually coincides with the exact energy that is conserved in gsBOMD.
The situation is different in CPMD. There, we also have time
translation invariance but the constant of motion reads
\begin{equation}
\label{eq:E_CP}
E_{CP}=\int\mathrm{d}\vec{r}\sum_A\frac{1}{2}
\mu_\mathrm{CP}\dot{\varphi}_A^*(\vec{r},t)\dot{\varphi}_A(\vec{r},t)+E\ ,
\end{equation}
which depends directly on the unphysical `mass' of the electrons,
$\mu_\mathrm{CP}$, and its conservation implies that the physical
energy $E$ varies in time. Still, this drawback has a minor effect,
since it has been shown that the CP physical energy follows closely
the exact gsBOMD energy curve.

The second global symmetry we want to consider is the change of orthonormal
basis of the space spanned by $\{\varphi_A\}_{A=1}^{n/2}$. Namely, given a
Hermitian matrix $S_{AB}$ ($S^+=S$), we define the following
transformation:
\begin{equation}
\label{eq:symmetry1}
\varphi^\prime_A=\sum_B\left(\mathrm{e}^{-iS}\right)_{AB}\varphi_B \ .
\end{equation}

The Lagrangian in~(\ref{eq:new_L}) depends on $\varphi$ only through
$\rho = 2 \sum_A | \varphi_A |^2$ and $\sum_A
\varphi_A^*\dot{\varphi}_A$. Provided the matrix \(S\) is Hermitian and
constant in time, both expressions are left unchanged by the
transformation. Hence, we can invoke again Noether theorem to obtain a
new conserved quantity that reads
\begin{equation}
\label{eq:conserved}
-i\sum_{A,B}\int \mathrm{d}\vec{r} 
\left[\frac{\delta L}{\delta\dot{\varphi}_A(\vec{r},t)}
S_{AB}\varphi_B(\vec{r},t)
- \frac{\delta L}{\delta\dot{\varphi}^*_A(\vec{r},t)}
S_{AB}\varphi^*_B(\vec{r},t)\right] = \mu\sum_{A,B} \int\,\mathrm{d}\vec{r} 
\varphi^*_A(\vec{r},t) S_{AB}\varphi_B(\vec{r},t) \ .
\end{equation}
Observe that we have a constant of motion for any Hermitian matrix
$S$. This permits us to combine different choices of \(S\) in order to
obtain that
\begin{equation}
\label{eq:escpro}
\int \mathrm{d}\vec{r} \varphi^*_A(\vec{r},t)\varphi_B(\vec{r},t)=\mathrm{const.}\ , \qquad \forall
  A,B=1,\ldots,\frac{n}{2} \ .
\end{equation}

In other words, if we start with an orthonormal set of wave functions
$\varphi$ and we let evolve the system according to
equations~(\ref{eq:new_EoM}), the family of wave functions maintains
its orthonormal character along time, i.e. the operator preserves the
inner product of the wave-functions that define the Ehrenfest
trajectory.

We would like to mention here that the above property is sometimes
substantiated on a supposed unitarity of the evolution
operator. \cite{Mar2000TR,The1992PRB,Kal1990IJSA}. Simply noticing
that the evolution of $\varphi$ is not linear as
both~(\ref{eq:new_EoM_a}) and~(\ref{eq:new_EoM_b}) are non-linear
equations, and that unitary evolution requires linearity \footnote{For
  a one to one transformation \(U\) we may define unitarity as the
  property of preserving the scalar product, i. e. given two arbitrary
  vectors $f$ and $g$ we have $\langle f, g \rangle =\langle Uf,
  Ug\rangle$.

  One can easily see that any reversible transformation $U$ that
  enjoys the previous property is necessarily linear:
  \begin{eqnarray*}
    \langle U(f_1+f_2), Ug\rangle &=& \langle f_1+f_2, g\rangle \\
    &=& \langle f_1, g\rangle + \langle f_2, g\rangle \\ 
    &=& \langle Uf_1, Ug\rangle + \langle Uf_2, Ug\rangle \\
    &=& \langle Uf_1 + Uf_2, Ug\rangle\ .
  \end{eqnarray*}
  As \(U\) is reversible, \(Ug\) is an arbitrary vector, and therefore
  we must have
  \begin{equation*}
    U(f_1+f_2) = Uf_1 + Uf_2\ .
  \end{equation*}

  Hence as the equations of motion (\ref{eq:new_EoM_a})
  and~(\ref{eq:new_EoM_b}) are clearly non linear, the evolution is
  non linear too and consequently it can not be a unitary
  transformation as defined above. If we fix the value of the density
  for all times, then the operator will become linear and it will be
  unitary.}, one can discard from the start the unitarity argument.


There is however a delicate point here that is worth discussing. The
issue is that the $\mu \to 0$ limit of our dynamics should correspond
to gsBOMD in eqns.~(\ref{eq:EoM_BO}) (which includes Lagrange
multipliers to keep orthonormalization), but the Lagrangian
in~(\ref{eq:new_L}) does not contain any multipliers and in fact they
are unnecessary, as our evolution preserves the
orthonormalization. This may rise some doubts on the equivalence
between gsBOMD and the limit of vanishing $\mu$ of our dynamics.

To settle the issue we introduce the additional dynamical fields
$\Lambda := (\Lambda_{AB})$, corresponding to Lagrange multipliers, in
our Lagrangian in eq.~(\ref{eq:new_L}), i.e.,
\begin{equation}
\label{eq:multipliers}
\tilde{L}[\varphi,\dot{\varphi},R,\dot{R},\Lambda]=
L[\varphi,\dot{\varphi},R,\dot{R}] + \sum_{AB}\Lambda_{AB}\left(\int\,\mathrm{d}\vec{r}
\varphi_A^*(\vec{r},t)\varphi_B(\vec{r},t) - \delta_{AB}\right) \ .
\end{equation}
This modification has an important consequence: the global symmetry
in~(\ref{eq:symmetry1}) becomes a gauge one with time-dependent matrix
elements. Actually one can easily verify that $\tilde{L}$ is invariant
under
\begin{subequations}
\label{eq:symmetry2}
\begin{align}
\varphi^\prime    & =  \mathrm{e}^{-iS}\varphi \ , \label{eq:symmetry2_a} \\
\Lambda^\prime & = \mathrm{e}^{-iS}\Lambda\mathrm{e}^{iS}
 - i\mu\mathrm{e}^{-iS} \frac{\mathrm{d}}{\mathrm{d}t}
   \mathrm{e}^{iS} \ . \label{eq:symmetry2_b}
\end{align}
\end{subequations}

This implies that, for $\mu \ne 0$, the fields $\Lambda_{AB}$ can be
transformed to any desired value by suitably choosing the gauge
parameters $S_{AB}(t)$. Their value is therefore irrelevant and one
could equally well take $\Lambda = 0$, as in (\ref{eq:new_L}), or
$\Lambda = \Lambda^\mathrm{BO}$ (the value it has in gsBOMD) without
affecting any physical observable. This solves the puzzle and shows
that the $\mu \to 0$ limit of the dynamics of eq.~(\ref{eq:new_EoM})
is in fact the exact gsBOMD.

\subsubsection{Physical interpretation}
\label{sec:phys}

If we take equation (\ref{eq:new_EoM_a}) and write the left hand side
as
\begin{equation}
  \label{eq:transf}
  \mu\frac{d\varphi}{dt} = \frac{d\varphi}{dt_e}\ , 
\end{equation}
the resulting equation can be seen as the standard Ehrenfest method in
terms of a fictitious time \(t_e\). Two important properties can be
obtained from this transformation.

On the one hand, it is easy to see that the effect of \(\mu\) is to
scale the TDDFT (\(\mu=1\)) excitation energies by a \(1/\mu\)
factor. So for \(\mu > 1\) the gap of the artificial system is
decreased with respect to the real one, while for small values of
\(\mu\), the excited states are pushed up in energy forcing the system
to stay in the adiabatic regime. This gives a physical explanation to
the \(\mu\to0\) limit shown before.

On the other hand, given the time step for standard Ehrenfest
dynamics, \(\Delta{}t\left(\mu=1\right)\), from (\ref{eq:transf}) we
can obtain that the time step as a function of \(\mu\) is
\begin{equation}
  \Delta{}t\left(\mu\right)=\mu\,\Delta{}t\left(\mu=1\right)\ ,
\end{equation}
so, for \(\mu>1\) propagation will be \(\mu\) times faster than
Ehrenfest.

By taking into account these two results we can see that there is a
trade-off in the value of \(\mu\): low values will give physical
accuracy while large values will produce a faster propagation. The
optimum value, that we will call \(\mu_{\mbox{max}}\), is the maximum
value of \(\mu\) that still keeps the system near the adiabatic
regime.  It is reasonable to expect that this value will be given by
the ratio between the electronic gap and the highest vibrational
frequency in the system. For many systems, like some molecules or
insulators, this ratio is large and we can expect large improvements
with respect to standard Ehrenfest MD. For other systems, like metals, this
ratio is small or zero and our method will not work well without
modifications (that are presently being worked on). We note that a
similar problem appears in the application of CP to these systems.

\subsection{Numerical properties}
\label{sec:numprop}

From the numerical point of view our method inherits the main
advantage of Ehrenfest dynamics: since propagation preserves
orthogonality of the wave-function, it needs not be imposed and the
numerical cost is proportional to \(N_{W}N_{C}\) (with \(N_{W}\) the
number of orbitals and \(N_{C}\) the number of grid points or basis
set coefficients). For CP a re-orthogonalization has to be done each
time step, so the cost is proportional to \(N_{W}^2N_{C}\). From these
scaling properties we can predict that for large enough systems our
method will be less costly than CP. As we will show below this
crossing can occur for around 1000 atoms for our implementation and
the systems we have considered.

Due to the complex nature of the propagator, Ehrenfest dynamics has to
be performed using complex wave-functions. In CP real wave-functions
can be used if the system is finite (without a magnetic field) or if
the system is a supercell using only the gamma point. However, with
respect to CP, the actual number of degrees of freedom to be treated
is the same, since CP equations are second order a second field has to
be stored, either the artificial ``velocity'' of the wave-functions or
the wave-function of the previous step.

An important point of comparison between Ehrenfest and CP is the
dependency of the maximum time step with the simulation parameters:
\(\mu\), \(\mu_{cp}\) and the cutoff energy (\(E_{cut}\)). While for
our modified Ehrenfest scheme it will scale like
\begin{equation}
  \label{eq:ehrenfesttimestep}
  \Delta{}t_{max}\propto\frac\mu{E_{cut}}\ ,
\end{equation}
for CP dynamics we have that\cite{Mar2000TR}
\begin{equation}
  \label{eq:cptimestep}
  \Delta{}t_{max}^{CP}\propto\sqrt{\frac{\mu_{cp}}{E_{cut}}}\ .
\end{equation}
Since \(\mu\) and \(\mu_{cp}\) are different quantities we cannot
infer anything without knowing the effect of their value in the
results, but as we will see from our calculations, even though in the
new scheme the time step increases linearly with \(\mu\), the
separation from the BO surface is also more sensitive to its value. On
the other hand, the dependence with the cutoff energy is one of the
major drawbacks of Ehrenfest dynamics, and probably it can explain
why, as we will see, it is slower than CP for small systems. However
in most cases this cutoff energy is independent from the size of the
system and will only represent a difference in the prefactor in the
scaling of both methods, so its effect should be compensated for large
systems.

\section{Methods}
\label{sec:Methods}

The scheme described above was implemented in the Octopus
code\cite{Octopus,Cas2006PSSB}. Octopus is general purpose code to
handle equilibrium and non-equilibrium phenomena using (TD)DFT.  It
can be used to simulate atoms, molecules, low dimensional systems, and
periodic structures under the presence of arbitrary electromagnetic
fields. The code is distributed under a free software license and many
new features are incorporated regularly. Octopus uses a real-space
grid representation combined with the finite differences approximation
for the calculation of derivatives\cite{Che1994PRL,
  Hirose2005Book}. The nuclei-electron interaction is replaced by
norm-conserving Troullier Martins pseudo-potentials. Unless stated
otherwise the Perdew-Zunger\cite{Per1981PRB} parametrization of the
Local Density Approximation (LDA) is used for the exchange and
correlation functional. The Poisson equation is solved using the
interpolating scaling functions method\cite{Gen2006JCP}.

\subsection{Time-propagation}

Given an initial condition \(\phi(t=0)\) and \(R(t=0)\), we want to
calculate \(\phi(t)\) and \(R(t)\) for a time \(t>0\) from
(\ref{eq:new_EoM}). For the ionic part, eq.~(\ref{eq:new_EoM_b}), once
the forces are computed\footnote{In principle the forces acting over
  the ions are given by eq.~(\ref{eq:new_EoM_b}), however, due to the
  derivatives of the ionic potential (that can have very high Fourier
  components), this expression is difficult to calculate accurately on
  real-space grids. Fortunately an alternative expression in terms of
  the gradient of the wave-functions can be obtained for both the
  local and non-local parts of the
  pseudo-potential\cite{Hirose2005Book}.}, the Newton equations can be
handled easily by the standard velocity Verlet algorithm.

For the electronic part, eq.~(\ref{eq:new_EoM_a}), the transformation
in eq.~(\ref{eq:transf}) allows us to use the standard Ehrenfest
propagation methods, making our scheme trivial to implement in an
existing real-time Ehrenfest code. The key part for the real-time
solution of equation~(\ref{eq:new_EoM_a}) is to approximate the
propagation operator
\begin{equation}
\varphi\left(t+\Delta t\right) 
= \hat{U}\left(t+\Delta t, t\right)\,\varphi\left(t\right)
\end{equation}
in an efficient and stable way. From the several methods available
(see ref.~\citealp{Cas2004JCP} for a review), in this work we have
chosen the {\it approximated enforced time-reversal symmetry} (AETRS)
method. For a Hamiltonian \(\hat{H}(t)\), in AETRS the propagator is
approximated by the explicitly time-reversible expression
\begin{equation}
\label{eq:aetrs}
\hat{U}\left(t+\Delta t, t\right)=
\exp\left\{-i\frac{\Delta{t}}2\hat{H}\left(t+\Delta{t}\right)\right\}
\exp\left\{-i\frac{\Delta{t}}2\hat{H}\left(t\right)\right\}\ ,
\end{equation}
with \(\hat{H}(t+\Delta{t})\) obtained from an interpolation from
previous steps. For the calculation of the exponential in
eq.~(\ref{eq:aetrs}) a simple fourth order Taylor expansion is
used. Note that the truncation to any order of the Taylor expansion
for the exponential operator implies that the norm of the vector is no
longer conserved. This theoretical error must be kept below an
acceptable threshold in order to ensure the preservation of the
orthonormality of the orbitals. In any case, a small inevitable error
will always lead to a slight change in the norm. If the norm is
reduced, the method is said to be ``contractive'' -- this property is
desirable since it leads to stable propagations, as opposed to the
case in which the norm increases: in this latter case the propagation
becomes unstable. The choice for a fourth order truncation is
advantageous because it is, for a very wide range of time-steps, a
contractive approximation to the exponential.

Moreover, the careful preservation of time reversibility is crucial to
avoid unphysical drifts in the total energy. We have found (for the
cases presented in this work and for our particular numerical
implementation) the combination of the AETRS approximation to the
propagator together with the Taylor expansion representation of the
exponential, to be the most efficient approach. Our tests show that
numerically the error in orthonormality, measured as the dot product
between orbitals, has an oscillatory behaviour and it is typically of
the order of \(10^{-10}\) but for some pairs of orbitals it can
increase to \(10^{-8}\).

We also implemented the Car-Parrinello scheme to compare it with our
approach. In this case the electronic part is integrated by the
RATTLE/Velocity Verlet algorithm as described in
ref.~\citealp{Tuc1994JCP}.

\subsection{Parallelization strategy}

The challenge of AIMD of going towards very large systems and large
simulation times is clearly linked to implementations that run
efficiently in parallel architectures. This is the case of CP methods,
that are known to perform very well in this kind of
systems\cite{Hut2005PC,Cav1999CPC}, the parallelization is usually based on
domain decomposition (known as parallelization over Fourier
coefficients in plane-wave codes) and K-points. However, good
scalability can only be obtained if the system is large enough to have
favorable computation-communication ratio with respect to the latency
of the interconnection.

This type of parallelization is also applicable to the present
Ehrenfest dynamics and, \emph{on top} of that, the new scheme can add
a different level of parallelization: since the propagation step is
independent for each orbital, it is natural to parallelize the problem
by distributing the Kohn-Sham states among processors. Communication
is only required once per time-step to calculate quantities that
depend on a sum over states: the time dependent densities and the
forces over the ions. This type of sum of a quantity over several
nodes is known as a reduction and the communication cost grows
logarithmically with the number of nodes.

The main limitation to the parallel scalability in our real space
implementation was observed to come from the parts of the code that do
not depend on the states (global quantities), mainly the re-generation
of the ionic potential
\begin{equation}
  \label{eq:ionicpotential}
  V^{ion}=\sum_J\hat{V}^{local}_{J}\left(\vec{r}-\vec{R_J}\right)
\end{equation}
and the calculation of the forces due to the local part of the ionic
potential
\begin{equation}
  \label{eq:ionicforces}
  \vec{F}^{local}_{J}=\int d\vec{r}\,
  \frac{d\rho\left(\vec{r}\right)}{d\vec{r}}\hat{V}^{local}_{J}\left(\vec{r}-\vec{R_J}\right)\ .
\end{equation}
As these expressions depend on the atoms index \(J\) a complementary
parallelization in atoms is used to speed-up these code sections. For
example, to generate the ionic potential, each processor generates the
potential for a subset of the atoms and then a reduction operation is
performed to obtain the total ionic potential.

Once this auxiliary parallelization over atoms is taken into account,
it results in a very efficient scheme, similar to K-point
parallelization for periodic systems, where, as long as there are
enough states to distribute, the scaling is linear even with slow
interconnections (as a rule of thumb, for our implementation 10-15
orbitals per processor are required for a good efficiency). In the
case of CP, due to orthogonalization between states the evolution is
not independent, so this parallelization scheme is more complex to
implement and requires more communication, making it much less
practical.

In our implementation we have combined this parallelization over
states with real space domain decomposition (see
ref.~\citealp{Cas2006PSSB} for details). This dual parallelization
strategy also has the advantage that allows us to decompose the two
levels of complexity, the size of the region of space simulated and
the number of orbitals, that increase when we move to study larger
systems.

Below we address the relative gain in performance of the code once
this second level of parallelization is used. To avoid as much as
possible issues related to different software packages we decided to
implement the two schemes, CP and Ehrenfest, in the same
code. Although this might not be the the best parallel implementation
of CP that is available in the community, it allows a direct
assessment of the impact of this extra level of parallelization. Given
the simplicity and the high level nature of parallelization over
states, it is expected that this gain will be transferable to other
implementations.

\section{Applications: model and realistic systems}
\label{sec:Numerical}

\subsection{Two band model}
\label{sec:Two_band}

To illustrate the properties of the new scheme, and also to compare it
to CP in a complementary manner to the calculations in the rest of the
manuscript, we apply it to a model system. The simple toy model we use
is based on the one used in the work by Pastore et
al. to test CP \cite{Pas1991PRA}. Its equations of motion are produced by the
Lagrangian
\begin{eqnarray}
\label{eq:L_toy}
L_{\mathrm{toy}} &=& \frac{\mu}{2} (\dot{\theta}_1 \theta_2 -
 \dot{\theta}_2 \theta_1 )
 + \frac{1}{2}M_R \dot{R}^2 + \frac{1}{2}M_G \dot{G}^2
\nonumber \\
&& \mbox{} - \frac{1}{2} K_R(R-R_0)^2
- \frac{1}{2} K_G(G-G_0)^2
 + \frac{G}{2}\left[\cos(\theta_1 - R) + \cos(\theta_2 - R)\right]
 \;,
\end{eqnarray}
where $\theta_1$ and $\theta_2$ correspond to electronic degrees of
freedom, $R$ to the nuclear motion and $G$ mimicks the gap. The
parameters $M_R$, $K_R$, $R_0$ and $G_0$ have been taken from the
experimental values for the N$_2$ molecule (interpreting $R$ as the
length of the N--N bond).

The dynamics produced by~(\ref{eq:L_toy}) has been then compared to the
analogous CP one [obtained by simply changing the $\theta$-kinetic
energy by $(\mu_\mathrm{CP}/2)(\dot{\theta}_1^2+\dot{\theta}_2^2)$],
and to the gsBO reference [defined by setting $\mu=0$, and $\theta_1$
and $\theta_2$ to the values that minimize the potential energy
in~(\ref{eq:L_toy}), $\theta_1=\theta_2=R$]. In all simulations, the
initial conditions of $R$ and $G$ have been increased a 10\% from
their equilibrium values $R_0$ and $G_0$, we have set
$\dot{R}(0)=\dot{G}(0)=0$, and the initial electronic coordinates have
been placed at the gsBO minimum (for CP,
$\dot{\theta}_1(0)=\dot{\theta}_2(0)=0$).

\begin{figure}
\begin{center}
\includegraphics[width=13cm]{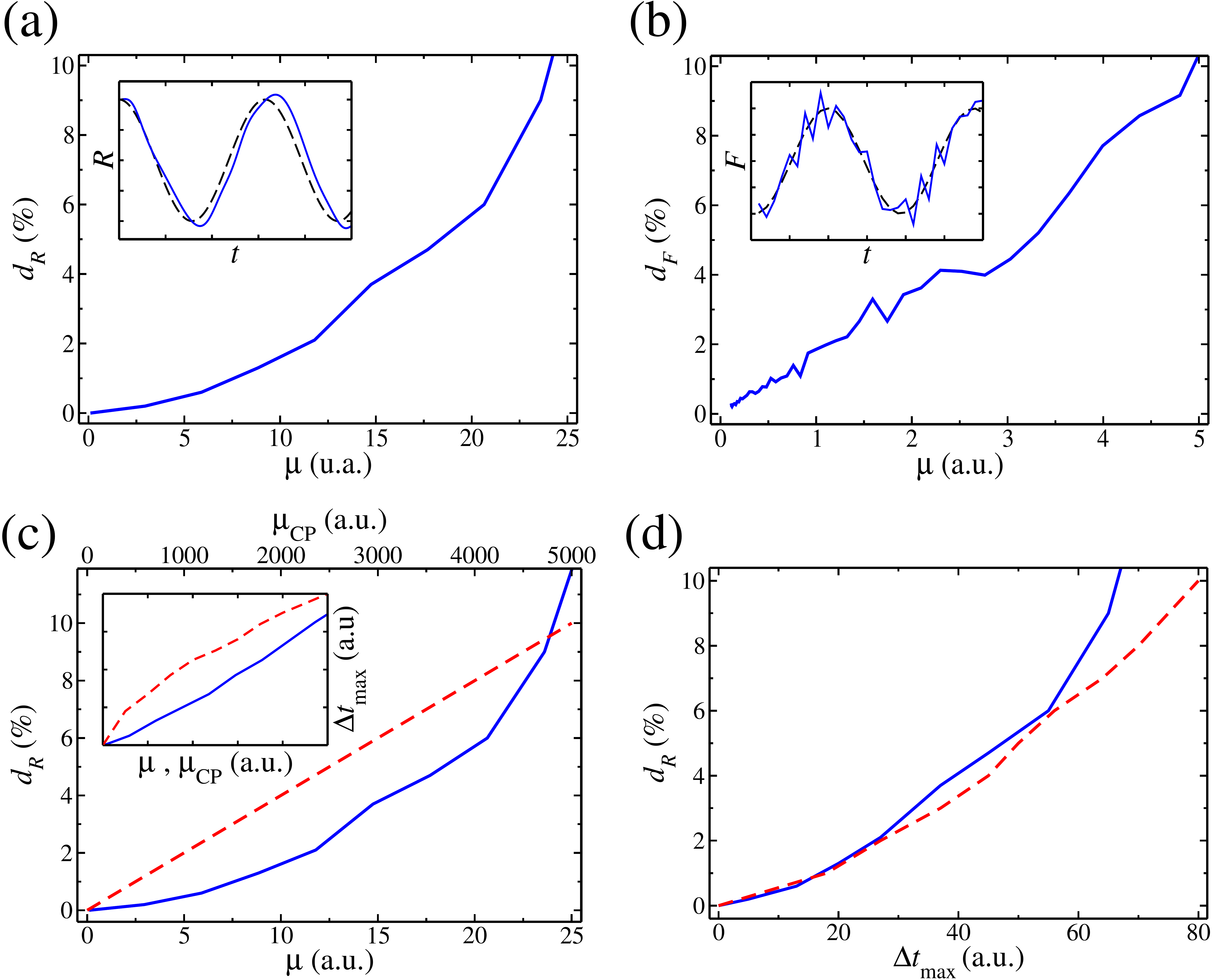}
\caption{\label{fig1} (a) Distance $d_R$ from the $R$-dynamics
  produced by~(\ref{eq:L_toy}) to the gsBO one as a function of $\mu$.
  Inset: the gsBO $R$-trajectory [broken (black) line] and the
  approximate one for $d_R=10\%$ [solid (blue) line]. (b) Distance
  between the averaged force on $R$ produced by~(\ref{eq:L_toy}) and
  the gsBO one. Inset: the gsBO force [broken (black) line] and the
  approximate one for $d_F=10\%$ [solid (blue) line]. (c) Dependence
  on $\mu$ (and $\mu_\mathrm{CP}$) of the distance $d_R$, and of the
  maximum time $\Delta t_\mathrm{max}$ step (inset); for both the new
  scheme [solid (blue) line] and CP [broken (red) line].  (d)
  Error/time step profile for both the new dynamics and CP [same keys
  as in (b)].}
\end{center}
\end{figure}

To compare the approximate nuclear trajectory $R(t)$ to the gsBO one
$R_\mathrm{BO}(t)$, we define $d_R := {100}/{\Delta R} \left(
  \frac{1}{T} \int_0^T \left[ R(t) - R_\mathrm{BO}(t) \right]^2
  \mathrm{d}t \right)^{1/2}$, where $\Delta R$ is the maximum
variation of $R$ in the gsBO case. In \ref{fig1}a, we show that this
distance smoothly decreases to zero as $\mu \to 0$ for our model. In
\ref{fig1}b, in turn, we compare the gsBO force on $R$ to the one
obtained from the new method averaging over a intermediate time
between those associated to the electronic and nuclear motions. The
distance $d_F$ between these forces (defined analogously to $d_R$),
also goes to zero when $\mu \to 0$.

Now, we estimate the relation between the maximum time step allowed by
the fourth order Runge-Kutta numerical integration of the equations of
motion and the error, given by $d_R$. The first, denoted by $\Delta
t_\mathrm{max}$, has been defined as the largest time step that
produced trajectories for all the dynamical variables of the system
with a distance less than 0.1\% to the `exact' trajectories. In
\ref{fig1}c, we can see that, although $\Delta t_\mathrm{max}$ grows
more slowly in our method than in CP (as expected from the discussion
in section \ref{sec:numprop}), the behaviour of the error ($d_R$) is
better for the new dynamics introduced here. These two effects
approximately balance each other yielding the error/time step
relations depicted in \ref{fig1}d, where the new scheme is shown to
behave similarly to CPMD for a significant range of values of
$d_R$. We stress however that, to actually compare the relative
performance of both methods the numerical work required in each time
step would have to be considered.  In this sense, the more realistic
simulations in the next sections are more representative.

\subsection{Nitrogen molecule}
\label{sec:Nitrogen}

\begin{figure}
\begin{center}
\includegraphics*[width=10cm]{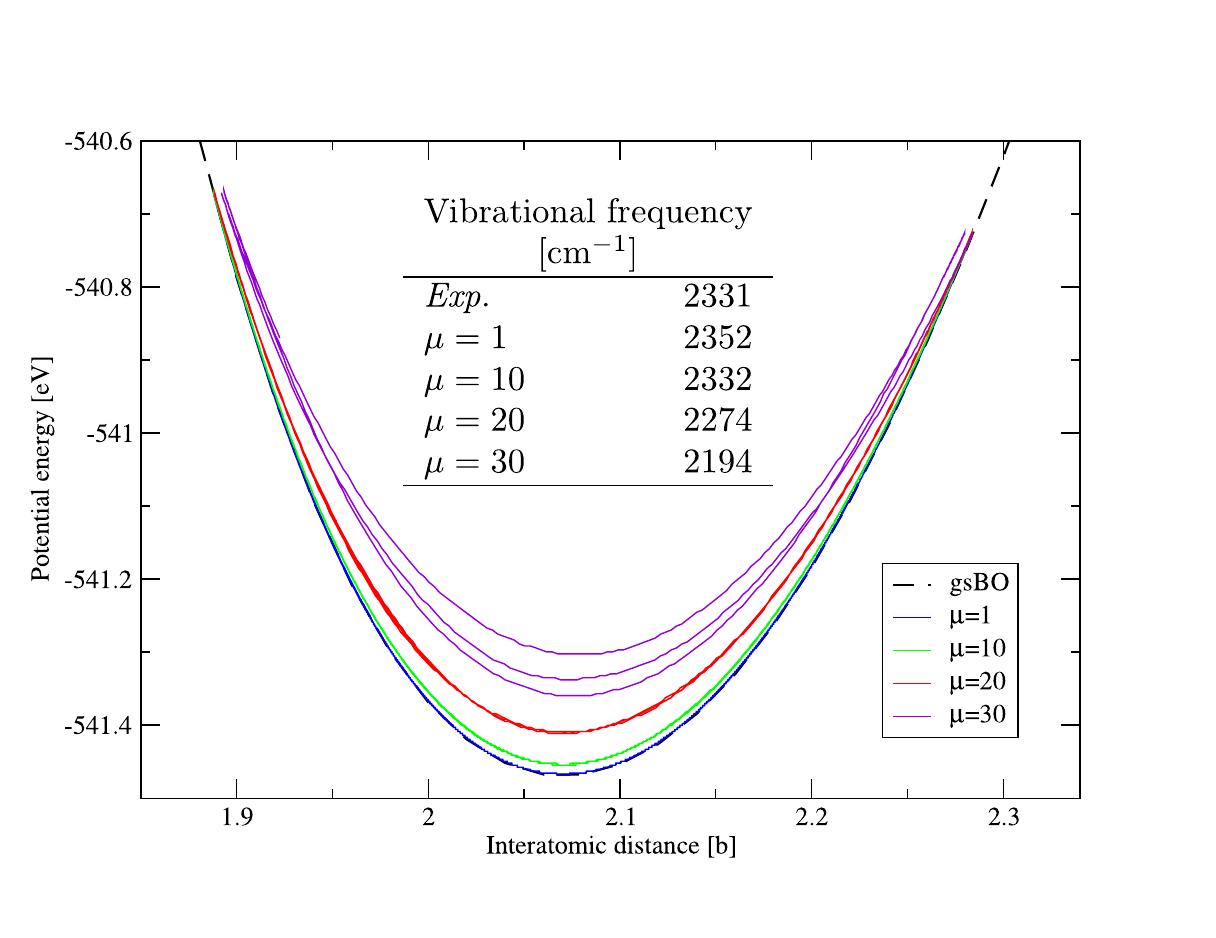}
\caption{\label{fig2} KS potential energy $E_\mathrm{KS}[\varphi,R]$
  as a function of the internuclear distance $R$ in N$_2$ molecule
  simulations. Starting from below, the gsBO result [broken (black)
  line], and $\mu=1$ [solid (blue) line], 10 [solid (green) line], 20
  [solid (red) line], 30 [solid (violet) line]. Inset: vibrational
  frequencies from experiment~\cite{Cra1949PR} and calculated from the
  trajectory using different values of $\mu$ and a spacing of 0.35 [b]
  and a box of radius 7.6 [b] around each atom.}
\end{center}
\end{figure}

For the Nitrogen molecule (N$_2$), we calculate the trajectories for
different values of $\mu$, using the same initial conditions as in the
toy model. A time step of $\mu \times 0.0012$ [fs] is used and the
system is propagated by $242$ [fs]. In \ref{fig2} we plot the
potential energy as a function of the interatomic distance during the
trajectory for each run, in the inset we also give the vibrational
frequency for the different values of $\mu$, obtained as the position
of the peak in the Fourier transform of the velocity auto-correlation
function. It is possible to see that for $\mu=20$ the simulation
remains steadily close to the BO potential energy surface and there is
only a $3.4\%$ deviation of the vibrational frequency. For $\mu=30$
the system starts to strongly separate from the gsBO surface as we
start to get strong mixing of the ground state BO surfaces with higher
energy BO surfaces. This behaviour is consistent with the physical
interpretation given in section~\ref{sec:phys} as for this system
\(\mu_{\mbox{max}}\approx27\).

\subsection{Benzene}
\label{sec:Benzene}

\begin{figure}[ht!]
\centering
\includegraphics*[width=\columnwidth]{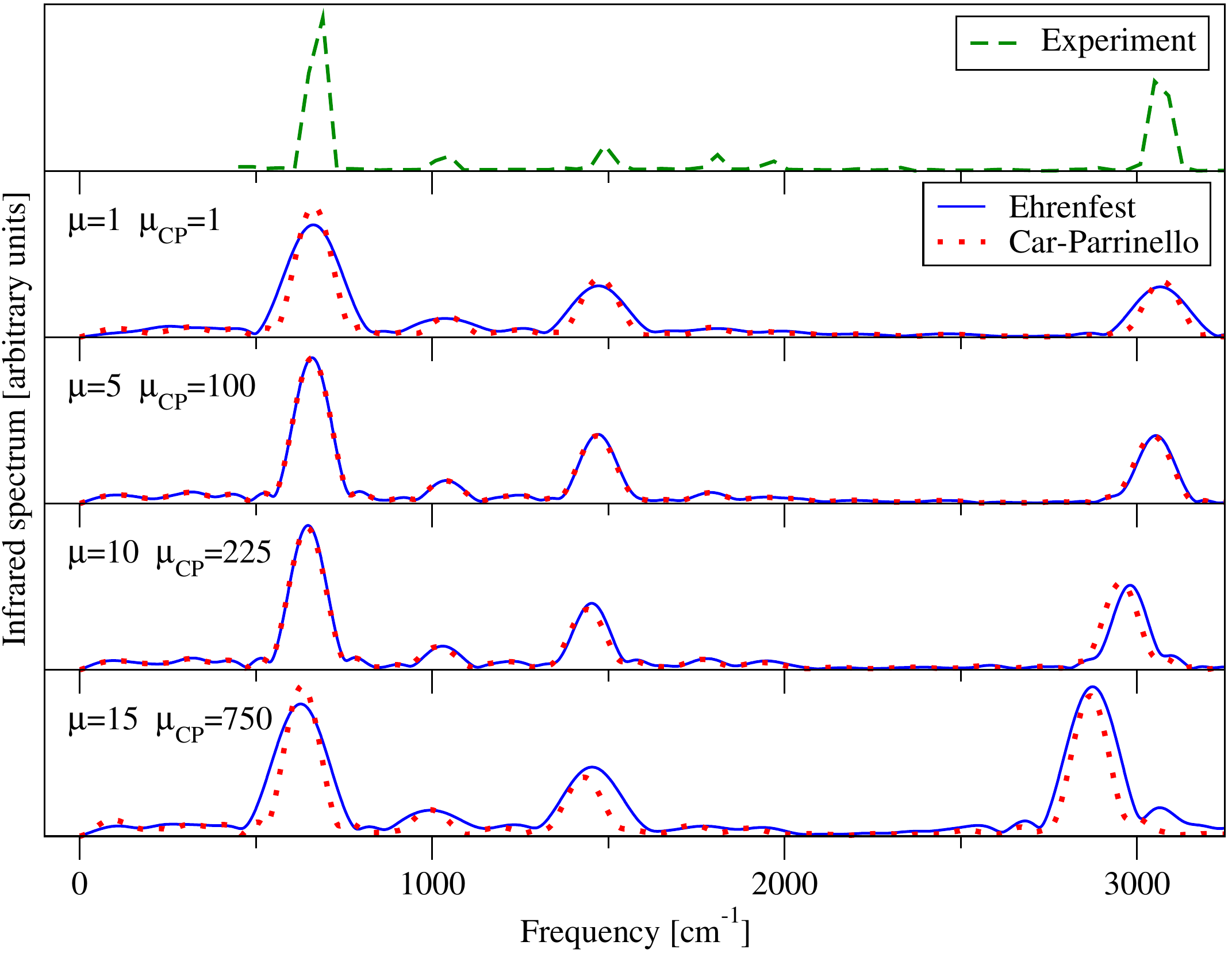}
\caption{\label{benzene} Calculated infrared spectrum for Benzene for
  different values of \(\mu\), compared to CP dynamics and to
  experiment from ref.~\citealp{benzene_experimental}. A spacing of
  0.35 [b] and a box of radius 7.6 [b] around each atom were used.}
\end{figure}

Next, we applied the method to the Benzene molecule. We set-up the
atoms in the equilibrium geometry with a random Maxwell-Boltzmann
distribution for \(300^\circ\mbox{K}\). Each run was propagated for a
period of time of \(\sim400\) [fs] with a time step of
\(\mu\times0.001\) [fs] (that provide a reasonable convergence in the
spectra). Vibrational frequencies were obtained from the Fourier
transform of the velocity auto-correlation function. In table
\ref{benz_freq}, we show some low, medium and high frequencies of
Benzene as a function of \(\mu\). The general trend is a red-shift of
the frequencies with a maximum deviation of 7\(\%\) for
\(\mu=15\). Still, to make a direct comparison with experiment, we
computed the infrared spectra as the Fourier transform of the
electronic dipole operator. In \ref{benzene}, we show how the spectra
changes with \(\mu\). For large \(\mu\), besides the red-shift,
spurious peaks appear above the higher vibrational frequency (not
shown), this is probably due to non-adiabatic effect as
\(\mu_{\mbox{max}}\approx14\) is we consider the first virtual TDDFT
excitation energy. We performed equivalent CP calculations for
different values of \(\mu_{cp}\), and found that, as shown in
\ref{benzene}, it is possible to relate the physical error induced in
both methods and establish and a relation between \(\mu\) and
\(\mu_{cp}\).

\begin{table}
\begin{tabular}{lrrrrrr}

\hline
\(\mu=1\)    & 398 & 961 & 1209 & 1623 & 3058 \\
\(\mu=5\)    & 396 & 958 & 1204 & 1620 & 3040 \\
\(\mu=10\)   & 391 & 928 & 1185 & 1611 & 2969 \\
\(\mu=15\)   & 381 & 938 & 1181 & 1597 & 2862 \\
\hline
\end{tabular}

\caption{\label{benz_freq} 
  Selected vibrational frequencies (in cm\(^{-1}\)) for the Benzene molecule,
  obtained using different values of \(\mu\) and a spacing of 0.35 [b]
  and a box of radius 7.6 [b] around each atom.
}
\end{table}

\begin{figure}[ht!]
\centering
\includegraphics*[width=\columnwidth]{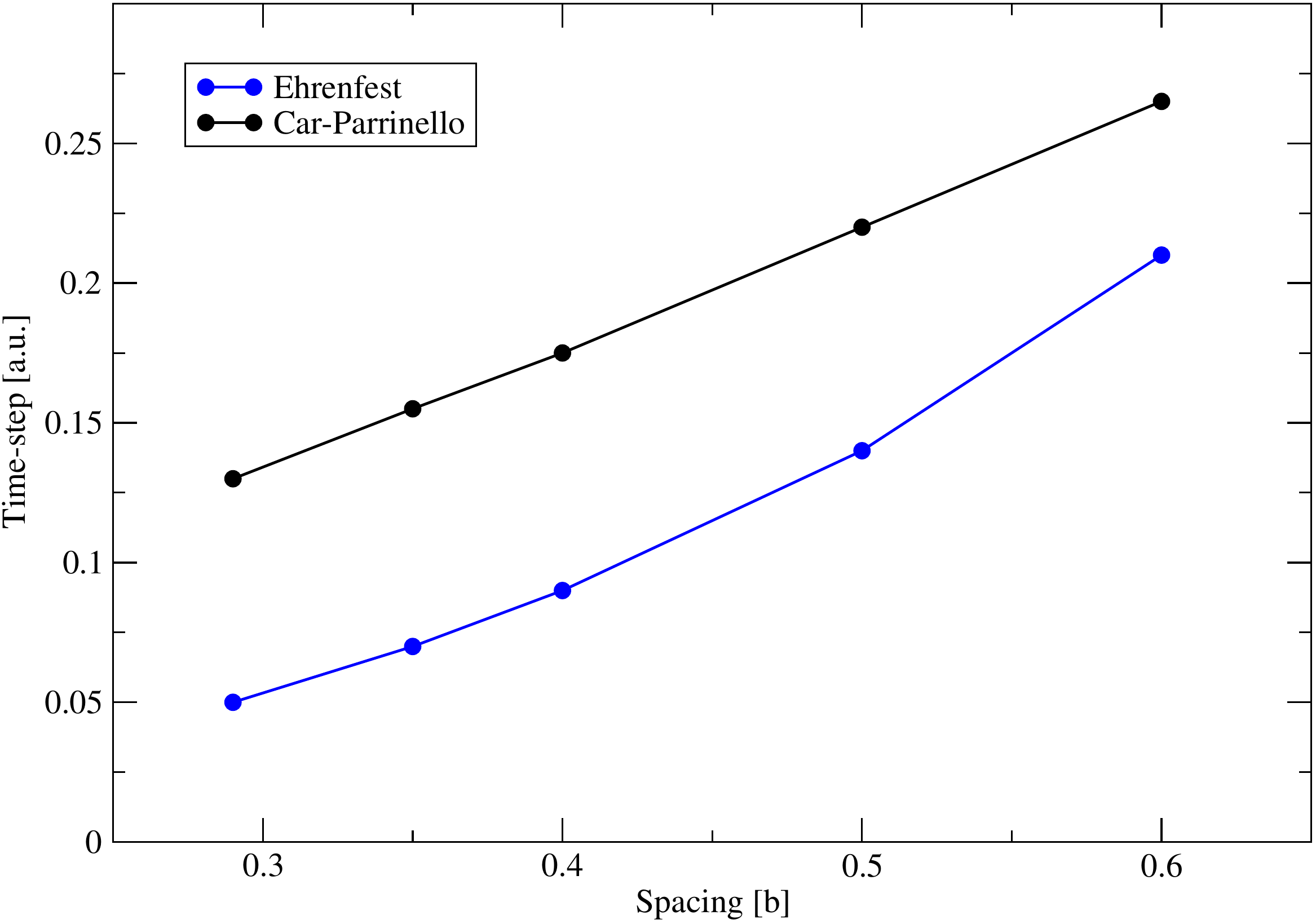}
\caption{\label{spacing} Comparison of the dependence of the time-step
  in terms of the spacing. In the case of Ehrenfest the time-step
  depends quadratically while in CP it is linear.}
\end{figure}

Having established the link between \(\mu\) and \(\mu_{cp}\) we
address the numerical performance of our new method compared to CP.
As explained in section~\ref{sec:numprop} the maximum time step has a
different behaviour with the cutoff energy or equivalently, in this
case, the grid spacing (the spacing is proportional to
\(\sqrt{1/E_{cut}}\)). This can be seen in \ref{spacing}, where we
plot the maximum time step both Ehrenfest and CP as a function of the
grid spacing. This is important since, in order to be able to do a
comparison for large number of atoms we use a larger spacing (0.6 [b]
or 14 [Ha]) than the required for the converged results previously
shown (0.35 [b] or 40 [Ha]). So for the small spacing case Ehrenfest
results should be scaled by a 1.7 factor, these two values gives us a
range of comparison, since most calculations are performed in this
range of precision (14 [Ha] to 80 [Ha]).

\begin{figure}[ht!]
\centering
\includegraphics*[width=\columnwidth]{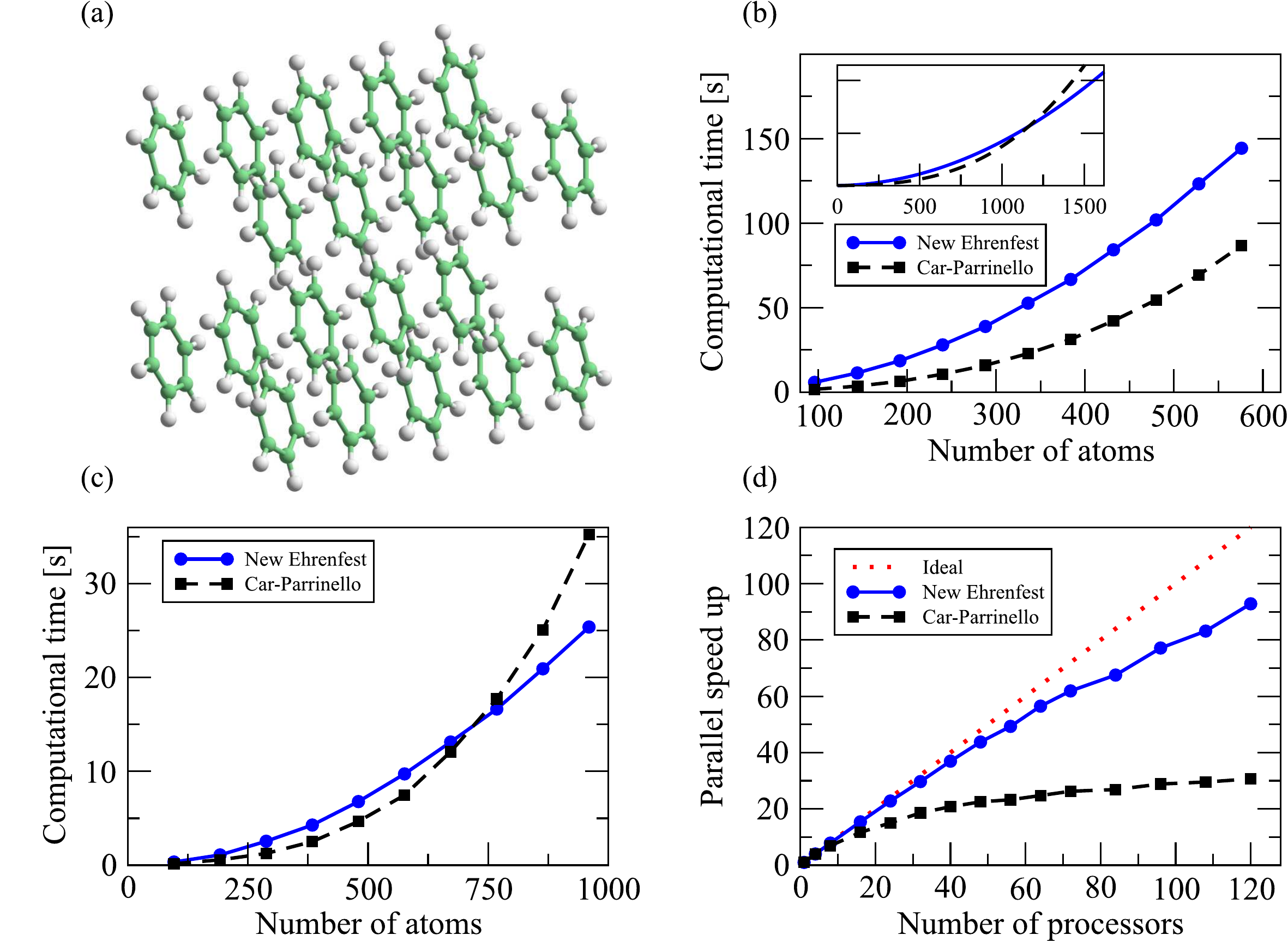}
\caption{\label{performance} (color online) Computational performance
  comparisons of our method, with \(\mu=15\) and CP, with
  \(\mu_{CP}=750\) for an array of Benzene molecules with finite
  boundary conditions and a spacing of 0.6 [b] and a box of radius 7.6
  [b] around each atom. Performance is measured as the computational
  time required to propagate one atomic unit of time.
  a) Scheme of the Benzene molecule array.
  b) Single processor computational cost for different system
  sizes. (inset) Polynomial extrapolation for larger
  systems. Performed in one core of an Intel Xeon E5435 processor.
  c) Parallel computational cost for different system sizes. Performed
  in \(32\,\times\,\)Intel Itanium 2 (1.66 GHz) processor cores of a
  SGI Altix.
  d) Parallel scaling with respect to the number of processor for a
  system of 480 atoms in a SGI Altix system. In both cases a mixed
  states-domain parallelization is used to maximize the performance.
}
\end{figure}

To compare in terms of system size, we simulated several Benzene
molecules in a cell. For the new scheme, a value of \(\mu=15\) is used
while for CP \(\mu_{cp}=750\), (values that yield a similar deviation
from the BO surface, according to \ref{benzene}). The time steps used
are \(3.15\) [a.u] and \(7.26\) [a.u.] respectively.  The
computational cost is measured as the simulation time required to
propagate one atomic unit of time, this is an objective measure to
compare different MD schemes. We performed the comparison both for
serial and parallel calculations; the results are shown in
\ref{performance}. In the serial case, CP is 3.5 times faster for the
smaller system, but the difference reduces to only 1.7 times faster
for the larger one. Extrapolating the results we predict that the new
dynamics will become less demanding than CP for around 1100 atoms, if
we consider the small spacing the crossover point moves to 2000
atoms. In the parallel case, the performance difference is reduced,
being CP only 2 times faster than our method for small systems, and
with a crossing point below 750 atoms (1150 atoms with the smaller
spacing). This is due to the better scalability of the Ehrenfest
approach, as seen on \ref{performance}c. Moreover, in our
implementation memory requirements as for our approach are lower than
for CP: in the case of 480 atoms the ground state calculation requires
a maximum of 3.5 Gb whereas in the molecular dynamics, Ehrenfest
requires \(5.6\) Gb while CP \(10.5\) Gb. The scaling of the memory
requirements is the same for both methods and we expect this
differences to remain proportional for all system sizes.

\subsection{Fullerene molecule: C\(_{60}\)}
\label{sec:Fullerene}

\begin{figure}[ht!]
\centering
\includegraphics*[width=\columnwidth]{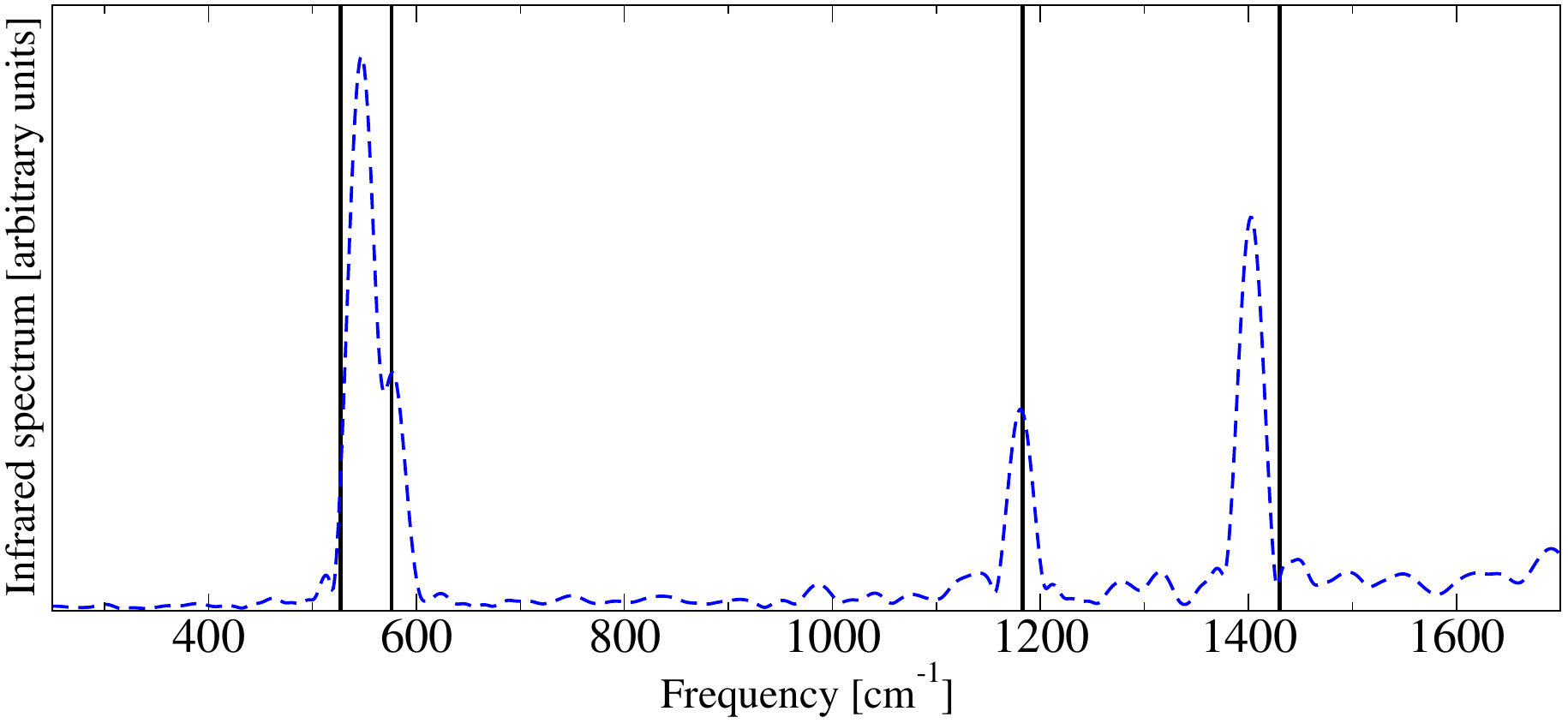}
\caption{\label{fullerenes} Infrared spectrum of C\(_{60}\). The
  (blue) dashed line corresponds to the calculated one (\(\mu\!=\!5\)
  and 2 [ps] of time) while the black bars are the experimental values
  from ref.~\citealp{Cab1998CPL}.  }
\end{figure}

To end the computational assessment of the new formalism, we
illustrate out method for the calculation of the infrared spectrum of
a prototype molecule, C\(_{60}\). This time we switch to the
PBE\cite{Per1996PRL} exchange and correlation functional since it
should give slightly better vibrational properties than
LDA\cite{Fav1999PRB}. For the simulation shown below we use a value of
\(\mu=5\) that provides a reasonable convergence in the spectra. The
calculated IR spectra is in very good agreement with the experiment
(see \ref{fullerenes}) for low and high energy peaks (the ones more
sensitive to the values of \(\mu\) as seen in \ref{benzene}). The
result is robust and independent of the initial condition of the
simulation. The low energy splitting of IR spectrum starts to be
resolved for simulations longer than 2 [ps].

\section{Conclusions}
\label{sec:Conclusions}

First principles molecular dynamics is usually performed in the
framework of ground-state Born-Oppenheimer calculations or
Car-Parrinello schemes. A major drawback of both methods is the
necessity to enforce the orthogonalization of the wave functions,
which can become the bottleneck for very large systems. Alternatively,
one can handle the electron-ion dynamics within the Ehrenfest scheme
where no explicit orthogonalization is necessary. However, in
Ehrenfest the time step needs to be much smaller than in both the
Born-Oppenheimer and the Car-Parrinello scheme. In this work we have
presented a new approach to AIMD based on a generalization of
Ehrenfest-TDDFT dynamics. This approach, we recall, relies on the
electron-nuclei separation ansatz, plus the classical limit for the
nuclei taken through short-wave asymptotics. Then, the electronic
subsystem is handled with time-dependent density-functional
theory. The resulting model consists of two coupled sets of equations:
the time-dependent Kohn-Sham equations for the electrons, and a set of
Newtonian equation for the nuclei, in which the expression for the
forces resembles, but is in fact unrelated to the
G{\"{u}}ttinger-Hellmann-Feynman form. We have stressed the relevance
of notational precision, in order to avoid this and other possible
common misunderstandings.

We shown how the new scheme preserves the desirable properties of
Ehrenfest allowing for a considerable increase of the time step while
keeping the system close to the Born-Oppenheimer surface. The
automatically enforced orthogonalization is of fundamental importance
for large systems because it reduces the scaling of the numerical cost
with the number of particles and, in addition, allows for a very
efficient parallelization, hence giving the method tremendous
potential for applications in computational science. Specially if the
method is integrated into codes that have other levels of
parallelization, enabling them to scale to even more processors or to
keeping the same level of parallel performance while treating smaller
systems.

Our approach introduces a parameter \(\mu\) that for particular values
recovers either Ehrenfest dynamics or Born-Oppenheimer dynamics. In
general \(\mu\) controls the trade-off between the closeness of the
simulation to the BO surface and the numerical cost of the calculation
(analogously to the role of the fictitious electronic mass in CP). We
have shown that for a certain range of values of \(\mu\) the dynamics
of the fictitious system is close enough to the Born-Oppenheimer
surface while allowing for a good numerical performance. We have made
direct comparisons of the numerical performance with CP, and, while
quantitatively our results are system- and implementation-dependent,
they prove that our method can outperform CP for some relevant
systems. Namely, large scale systems that are of interest in several
research areas and that can only be studied from first principles MD
in massively parallel computers. To increase its applicability it
would be important to study if the improvements developed to optimize
CP can be combined with our approach~\cite{Kuhne2007PRL}, in
particular techniques to treat small gap or metallic
systems~\cite{Marzari1997PRL}.

Note that the introduction of the parameter \(\mu\) comes at
a cost, as we change the time scale of the movements of the electrons
with respect to the Ehrenfest case, which implies a shift in the
electronic excitation energies. This must be taken into account to
extend the applicability of our method for non-adiabatic MD and MD
under electromagnetic fields, in particular for the case of Raman
spectroscopy, general resonant vibrational spectroscopy as well as
laser induced molecular bond rearrangement. In this respect,
we stress that in the examples presented in this work, we have
utilized the new model to perform Ehrenfest dynamics in the limit
where this model tends to ground state, adiabatic MD. In this case, 
as it became clear with these examples, the attempt to
gain computational performance by enlarging the value of the parameter
$\mu$ must be done carefully, since the non-adiabatic influence of the
higher lying electronic states states increases with increasing
$\mu$. We believe, however, that there are a number of avenues to be
explored that could reduce this undesired effect; we are currently
exploring the manner in which the "acceleration" parameter $\mu$ can
be introduced while keeping the electronic system more isolated from
the excited states.

Nevertheless, Ehrenfest dynamics incorporates in principle the
possibility of electronic excitations, and non-adiabaticity.  The
proper incorporation of the electronic response is crucial for
describing a host of dynamical processes, including laser-induced
chemistry, dynamics at metal or semiconductor surfaces, or electron
transfer in molecular, biological, interfacial, or electrochemical
systems. The two most widely used approaches to account for
non-adiabatic effects are the surface-hopping method and the Ehrenfest
method implemented here. The surface-hopping approach extends the
Born-Oppenheimer framework to the non-adiabatic regime by allowing
stochastic transitions subject to a time- and momenta-dependent
hopping probability. On the other hand Ehrenfest successfully adds
some non-adiabatic features to molecular dynamics but is rather
incomplete. This approximation can fail either when the nuclei have to
be treated as quantum particles (e.g. tunnelling) or when the nuclei
respond to the microscopic fluctuations in the electron charge density
(heating)\cite{Hor2006RPP} not reproducing the correct thermal
equilibrium between electrons and nuclei (which constitutes a
fundamental failure when simulating the vibrational relaxation of
biomolecules in solution).  We have briefly addressed these issues in
section~\ref{sec:Ehrenfest}; as mentioned there, there have been some
proposals in the literature to modify Ehrenfest in order to fulfill
Boltzmann
equilibrium\cite{Bastida2006,Bas2007JCP,Tully1990}. Currently we are
also investigating related extensions to Ehrenfest to obtain the
correct equilibrium in our simulations.

\section*{Acknowledgements}

\hspace{0.5cm} We would like to thank A. Bastida, G. Ciccotti and E.K.U.
Gross for illuminating discussions.

This work has been supported by the research projects DGA (Arag\'on
Government, Spain) E24/3 and MEC (Spain)
\mbox{FIS2006-12781-C02-01}. P. Echenique is supported by a MEC/MICINN
(Spain) postdoctoral contract. X.A and A.R. acknowledge funding by the
Spanish MEC (FIS2007-65702-C02-01), "Grupos Consolidados UPV/EHU del
Gobierno Vasco" (IT-319-07), and the European Community through NoE
Nanoquanta (NMP4-CT-2004-500198), e-I3 ETSF (INFRA-2007-1.2.2: Grant
Agreement Number 211956), NANO-ERA Chemistry, DNA-NANODEVICES
(IST-2006-029192) and SANES (NMP4-CT-2006-017310)
projects. Computational resources were provided by the Barcelona
Supercomputing Center, the Basque Country University UPV/EHU (SGIker
Arina) and ETSF.

{
\phantomsection
\addcontentsline{toc}{chapter}{References}
\bibliography{modEhrenfest}
}

\end{document}